\documentclass[%
 preprint,
superscriptaddress,
 amsmath,amssymb,
 aps,
prb,
]{revtex4-2}

\usepackage{graphicx}
\usepackage{dcolumn}
\usepackage{bm}
\usepackage[version = 4]{mhchem}
\usepackage{soul, xcolor}
\usepackage{sidecap}
\usepackage{multirow}
\usepackage{makecell}
\usepackage{comment}
\usepackage{orcidlink}
\usepackage[normalem]{ulem}
\usepackage{booktabs}
\usepackage{placeins}
\usepackage{float}

\begin{document}
\title{
\texorpdfstring{
Transforming Native Oxide into a Metallic Platinum--Niobium Alloy for Passivation of Superconducting Niobium Films
}{
Transforming Native Oxide into a Metallic Platinum-Niobium Alloy for Passivation of Superconducting Niobium Films
}
}

\author{Ananya Chattaraj\,\orcidlink{0000-0002-9225-5224}}
\affiliation{Center for Functional Nanomaterials, Brookhaven National Laboratory, Upton, NY 11973, USA.}%

\author{Conan Weiland\,\orcidlink{0000-0001-6808-1941}}
\affiliation{Material Measurement Laboratory, National Institute of Standards and Technology, Gaithersburg, MD, 20899 USA}

\author{Bruce Ravel\,\orcidlink{0000-0002-4126-872X}}
\affiliation{Material Measurement Laboratory, National Institute of Standards and Technology, Gaithersburg, MD, 20899 USA}

\author{Kim Kisslinger\,\orcidlink{0000-0002-6528-7044}}
\affiliation{Center for Functional Nanomaterials, Brookhaven National Laboratory, Upton, NY 11973, USA.}%

\author{Sooyeon Hwang\,\orcidlink{0000-0001-5606-6728}}
\affiliation{Center for Functional Nanomaterials, Brookhaven National Laboratory, Upton, NY 11973, USA.}%

\author{Ming Lu\,\orcidlink{0000-0003-0175-6531}}
\affiliation{Center for Functional Nanomaterials, Brookhaven National Laboratory, Upton, NY 11973, USA.}%

\author{Nikhil Tiwale\,\orcidlink{0000-0001-8229-7108}}
\affiliation{Center for Functional Nanomaterials, Brookhaven National Laboratory, Upton, NY 11973, USA.}%

\author{Xiao Tong\,\orcidlink{0000-0002-5567-9677}}
\affiliation{Center for Functional Nanomaterials, Brookhaven National Laboratory, Upton, NY 11973, USA.}%

\author{Ajith Pattammattel\,\orcidlink{0000-0002-5956-7808}}
\affiliation{National Synchrotron Light Source II, Brookhaven National Laboratory, Upton, NY 11973, USA.}%

\author{Andrew M. Kiss\,\orcidlink{0000-0002-8515-5508}}
\affiliation{National Synchrotron Light Source II, Brookhaven National Laboratory, Upton, NY 11973, USA.}%

\author{Steven L. Hulbert\,\orcidlink{0000-0003-3127-6029}}
\affiliation{National Synchrotron Light Source II, Brookhaven National Laboratory, Upton, NY 11973, USA.}%

\author{Aswin kumar Anbalagan\,\orcidlink{0000-0001-5511-2083}}
\email{Contact author: aanbalaga1@bnl.gov}
\affiliation{National Synchrotron Light Source II, Brookhaven National Laboratory, Upton, NY 11973, USA.}%

\author{Andrew L. Walter\,\orcidlink{0009-0002-4215-8546}}
\email{Contact author: awalter@bnl.gov}
\affiliation{National Synchrotron Light Source II, Brookhaven National Laboratory, Upton, NY 11973, USA.}%

\author{Peter V. Sushko\,\orcidlink{0000-0001-7338-4146}}
\email{Contact author: peter.sushko@pnnl.gov}
\affiliation{Integrated Discovery Sciences Directorate, Pacific Northwest National Laboratory, Richland, WA 99354, USA.}%

\author{Mingzhao Liu\,\orcidlink{0000-0002-0999-5214}}
 \email{Contact author: mzliu@bnl.gov}
\affiliation{Center for Functional Nanomaterials, Brookhaven National Laboratory, Upton, NY 11973, USA.}%

\date{\today}

\begin{abstract}

Dielectric losses at surfaces and interfaces remains a primary limitation to coherence in superconducting transmon qubits. Niobium (Nb), a widely used material in superconducting quantum circuits, readily forms native oxides under ambient conditions, leading to lossy dielectric interfaces that degrade device performance. Here, we demonstrate a novel fabrication strategy that chemically transforms the native dielectric oxide on Nb into a metallic Pt--Nb surface alloy through post-oxidation Pt deposition followed by thermal annealing, thereby stabilizing Nb surfaces against further oxidation, potentially including sidewall regions. High-purity Nb films were fabricated with bulk-like superconducting transition temperatures ($T_c = 9.30\pm0.10$~K). Spectroscopic and microscopic analyses confirm the conversion of the native oxide into a chemically stable, metallic alloy-dominated layer that suppresses subsequent oxide growth. Ab initio simulations elucidate the atomic-scale rearrangement and electronic structure evolution associated with Pt incorporation into native niobium oxide, providing insight into the mechanism underlying the oxide-to-metal transformation and stabilization of the alloyed surface.

\end{abstract}

\maketitle

\newpage


\section*{Introduction}
Superconducting transmon qubits provide one of the leading platforms for quantum computing, exploiting nonlinear inductance from Josephson junctions to store and manipulate quantum information. Their appeal stems from relatively long coherence times, scalability, and compatibility with established microfabrication techniques. However, their performance remains fundamentally limited by dielectric losses, predominantly due to two-level systems (TLS) found at surfaces and interfaces \cite{MARTINIS2004487,Kjaergaard2020SuperconductingQubits,Wang2015SurfaceParticipation,Dial2016BulkSurfaceLoss,Place2021NewMaterialPlatform,chattaraj2025nbfilms}.

Niobium (Nb, $T_c = 9.3$ K) and tantalum (Ta, $T_c = 4.3$ K) are widely used as superconducting materials in transmon qubits \cite{Place2021NewMaterialPlatform, chattaraj2025nbfilms, anferov2024niobium}.
Despite its higher $T_c$, the performance of Nb-based devices is consistently limited by the rapid formation of native oxides (e.g., \ce{Nb2O5}, \ce{NbO2}, \ce{NbO}), which introduce dielectric losses and degrade device coherence \cite{Murthy2022NbOxideQubit,Kalboussi2025NbOxidesTLS,Berman2023Nb110Oxygen,Verjauw2021NbResonatorLoss}. Premkumar \textit{et al.}\ showed that microscopic relaxation pathways in superconducting materials are strongly influenced by oxide stoichiometry and suboxide penetration \cite{Premkumar2021Microscopic}. Altoé \textit{et al.}\ further demonstrated that process-induced oxides and fabrication residues constitute dominant TLS channels that persist despite standard cleaning procedures, highlighting the need for more robust passivation strategies \cite{Altoe2022Localization}. 

Efforts to mitigate dielectric losses in Nb-based devices have included surface processing approaches aiming to remove or transform the native surface oxides, such as vacuum annealing and wet chemical etching (e.g., HF treatments). However, these approaches do not ensure a long-term stability of the Nb surfaces as the oxide layer regrows rapidly under ambient conditions \cite{Kopas2024AmmoniumFluorideQubit,TorresCastanedo2024NbHydrides,Earnest2018SiAlResonators}. Pursuing substitution of native surface oxide with alternatives using more advanced methods, such as nitrogen plasma treatments \cite{Zheng2022Nitrogen}, coating with self-assembled monolayers (SAMs) \cite{Alghadeer2024Mitigating}, and engineered oxides formed via neutral-atom beam processing \cite{Kar2023NeutralAtomBeam}, have demonstrated partial improvements. However, these approaches do not eliminate the presence of surface oxides and, in many cases, leave behind dielectric layers whose thickness, composition, and defect content evolve over time.

Several groups have explored encapsulation strategies to suppress native oxide formation in Nb- and Ta-based qubits, using materials that either resist oxidation or form less lossy native oxides. For example, de Ory \textit{et al.}\ demonstrated low-loss Nb/Au hybrid resonators \cite{CaleroDeOry2025LowLoss}, while Bal \textit{et al.}\ reported significant improvements in transmon coherence through Nb surface encapsulation \cite{Bal2024Systematic}. In most implementations, the encapsulation layer is deposited \emph{in situ} immediately after deposition of the superconducting film, thereby preventing oxidation of the surface. However, as noted by Chang \textit{et al.} \cite{Chang2025SurfaceOxide}, subsequent device patterning inevitably exposes sidewall regions that remain unprotected and are therefore susceptible to oxidation. These sidewalls are particularly detrimental, as they typically exhibit higher electric-field participation and increased surface roughness. In other words, the \emph{in situ} encapsulation method cannot prevent sidewall oxidation, thus greatly limiting the overall effectiveness of this strategy.

In this work, we report a novel encapsulation strategy that addresses the otherwise unavoidable oxidation of device sidewalls. In this approach, a thin platinum (Pt) layer is deposited onto Nb that has been exposed to air and developed a native oxide (NbO$_x$). Subsequent annealing drives oxygen diffusion into the Nb film, effectively reducing the surface oxide while promoting the formation of a Nb–Pt alloy. As a result, the initially oxide-covered surface is transformed into a metallic alloy-dominated layer that is stable against further oxidation, making this approach promising for sidewall passivation. We investigate this encapsulation process using X-ray diffraction (XRD), transmission electron microscopy (TEM), scanning transmission electron microscopy (STEM), variable-energy X-ray photoelectron spectroscopy (VEXPS), and extended X-ray absorption fine structure (EXAFS). In addition, density functional theory (DFT) calculations provide a mechanistic picture of the atomic rearrangements driven by Pt deposition that results in the dissolution of the oxide and formation of thermodynamically stable Nb–Pt alloy.

\section*{Results and discussion}

\begin{figure}
\includegraphics[width=1.0\columnwidth]{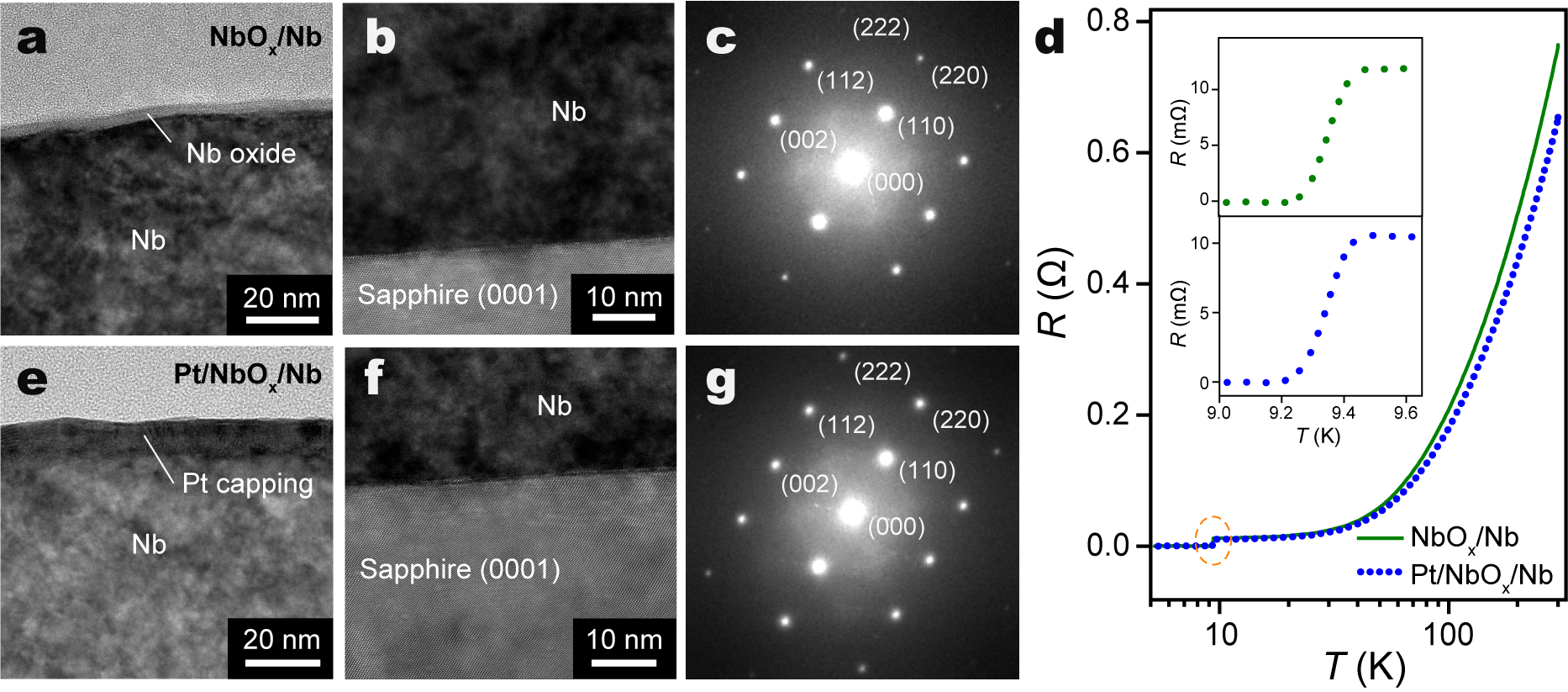}

\caption{(a, b) HRTEM images of bare Nb film near its top surface and near the Nb/\ce{Al2O3} interface. (c) Fast Fourier transform (FFT) obtained from the bulk Nb region marked in (c), indexed to bcc Nb with a zone axis of $[\bar{1}10]$. (d) $R-T$ relations of NbO$_x$/Nb (green solid line) and Pt/NbO$_x$/Nb (blue dotted line) thin films, showing superconducting transitions at $T_c\approx9.3$ K (insets). (e, f) HRTEM images of Pt/NbO$_x$/Nb film near its top surface and near the Nb/\ce{Al2O3} interface. (g) FFT obtained from the bulk Nb region marked in (f), indexed to bcc Nb with a zone axis of $[\bar{1}10]$.}
\label{fig1:TEM-RT}
\end{figure}
\begin{figure}
\includegraphics[width=1.0\columnwidth]{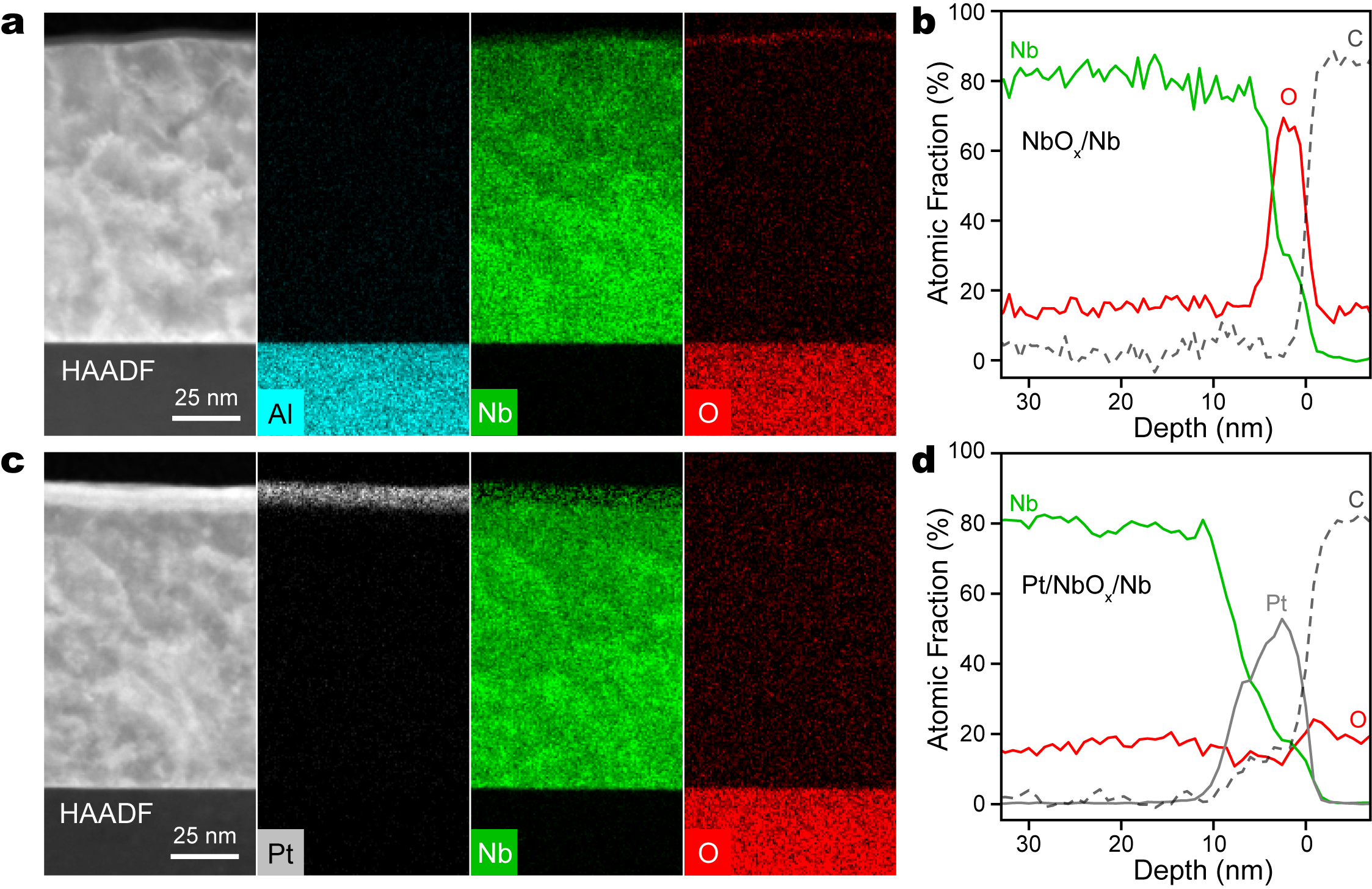}

\caption{(a) Cross-sectional HAADF-STEM image of a bare Nb thin film deposited on \ce{Al2O3}, and corresponding EDX elemental maps of Al (cyan), Nb (green), and O (red), respectively. (b) Integrated elemental distribution of Nb (green) and O (red) near the top surface of the bare Nb film. Depth is measured from the top surface of native oxide. (c) Cross-sectional HAADF-STEM image of a bare Nb thin film deposited on \ce{Al2O3}, and corresponding EDX elemental maps of Pt (gray), Nb (green), and O (red), respectively. (d) Integrated elemental distribution of Pt (gray), Nb (green) and O (red) near the top surface of the bare Nb film. Depth is measured from the top surface of native oxide. In both (b) and (d), amorphous carbon is deposited on the film surface during FIB liftoff and considered in the elemental analysis (dashed gray).}
\label{fig2:STEM-EDX}
\end{figure}

 Nb films are deposited on c-plane sapphire substrates using magnetron sputtering, at a substrate temperature of 700 $^\circ$C, to a typical thickness of $\approx$ 110 nm. According to its XRD pattern, the Nb film adopts the bulk body-center cubic (bcc) lattice with a preferred (111)-orientation, consistent with epitaxial growth on c-plane sapphire (Supporting Information, Figure S1). High-resolution TEM and selected-area electron diffraction (SAED) confirm the Nb (111)$\parallel$\ce{Al2O3} (0001) epitaxial relation, showing that the Nb atomic lattice aligns coherently with the sapphire substrate (Figure \ref{fig1:TEM-RT}a-c). Cross-sectional high-angle annular dark-field (HAADF) STEM image and energy-dispersive X-ray spectroscopy (EDX) mapping reveal sharp Nb/sapphire interfaces and uniform elemental distributions across all layers (Figure \ref{fig2:STEM-EDX}). Unlike the columnar microstructure observed in Nb films deposited at room temperature \cite{Chattaraj2025NbVacuum}, the high-temperature-grown Nb film grown in this work is single crystalline and free of structural defects, such as grain boundaries and amorphous regions. Temperature-dependent electrical resistance measurement on as-deposited Nb thin film reveals exceptionally high material quality. The film exhibits a superconducting transition temperature $T_c = (9.30\pm0.10$) K, identical to that of the bulk Nb, with a narrow transition width that suggests a spatially uniform superconducting order parameter (Figure \ref{fig1:TEM-RT}d). The residual resistance ratio (RRR), defined as $R_\mathrm{300 K}/R_\mathrm{10 K}$, exceeds 63, further demonstrating its bulk-like transport properties.\cite{chattaraj2025nbfilms}

After the Nb film is removed from the deposition chamber and exposed to ambient air about an hour, a self-limiting native oxide layer forms, which has an average thickness of ($2.6\pm0.2$) nm according to the TEM imaging (Figures \ref{fig1:TEM-RT}, \ref{fig2:STEM-EDX}a, and Figure S2c). In the discussion that follows, the native oxide-covered Nb films will be referred to as \ce{NbO_x}/Nb. In actual device fabrication, although the Nb oxide thickness can vary with processing history, fluoride-based chemical treatments such as buffered oxide etch can readily reduce it to 1--3~nm\cite{Altoe2022Localization}, making the oxide studied here a technologically relevant model for processed Nb surfaces.

The \ce{NbO_x}/Nb film was subsequently returned to the deposition chamber for deposition of an approximately 10~nm thick Pt overlayer by sputtering (details in Methods), followed by vacuum annealing at 750~$^\circ$C. The annealing duration was varied from 0.5 to 2.0~h under otherwise identical conditions. As the interfacial evolution approached a near-steady state after approximately 1~h, a duration of 2.0~h was adopted for the detailed multimodal characterization described below. TEM imaging suggests that the encapsulation layer forms a continuous, uniform cap without observable voids or delamination (Figure~\ref{fig2:STEM-EDX}c). In contrast to elemental Pt, the encapsulation is resistant to chemical etchant such as \textit{aqua regia}, suggesting the formation of intermetallic alloys. Accordingly, should this process be adapted for actual device fabrication, the difference in chemical properties may enable the selective removal of unreacted Pt for self-aligned encapsulation of Nb.

On the other hand, the native oxide layer that covered the Nb film prior to the Pt coating is no longer detectable in the EDX mapping, suggesting that the oxygen migration takes place during the Pt deposition and the vacuum annealing that follows (Figures \ref{fig2:STEM-EDX}c, d). The encapsulation and subsequent annealing have no apparent effect on the bulk of Nb film, which remains single crystalline (Figures \ref{fig1:TEM-RT}e-g). As a result, the electrical transport properties are very similar to those of the as-grown Nb film (\ce{NbO_x}/Nb), with virtually identical $T_c$ ($9.30\pm0.10$) ~K and RRR (64) (Figure \ref{fig1:TEM-RT}d). The observation that the encapsulation layer does not degrade the DC superconducting transport metrics suggests that the process has no negative impact on the bulk region of the Nb film.

\begin{figure}
\includegraphics[width=1.0\columnwidth]{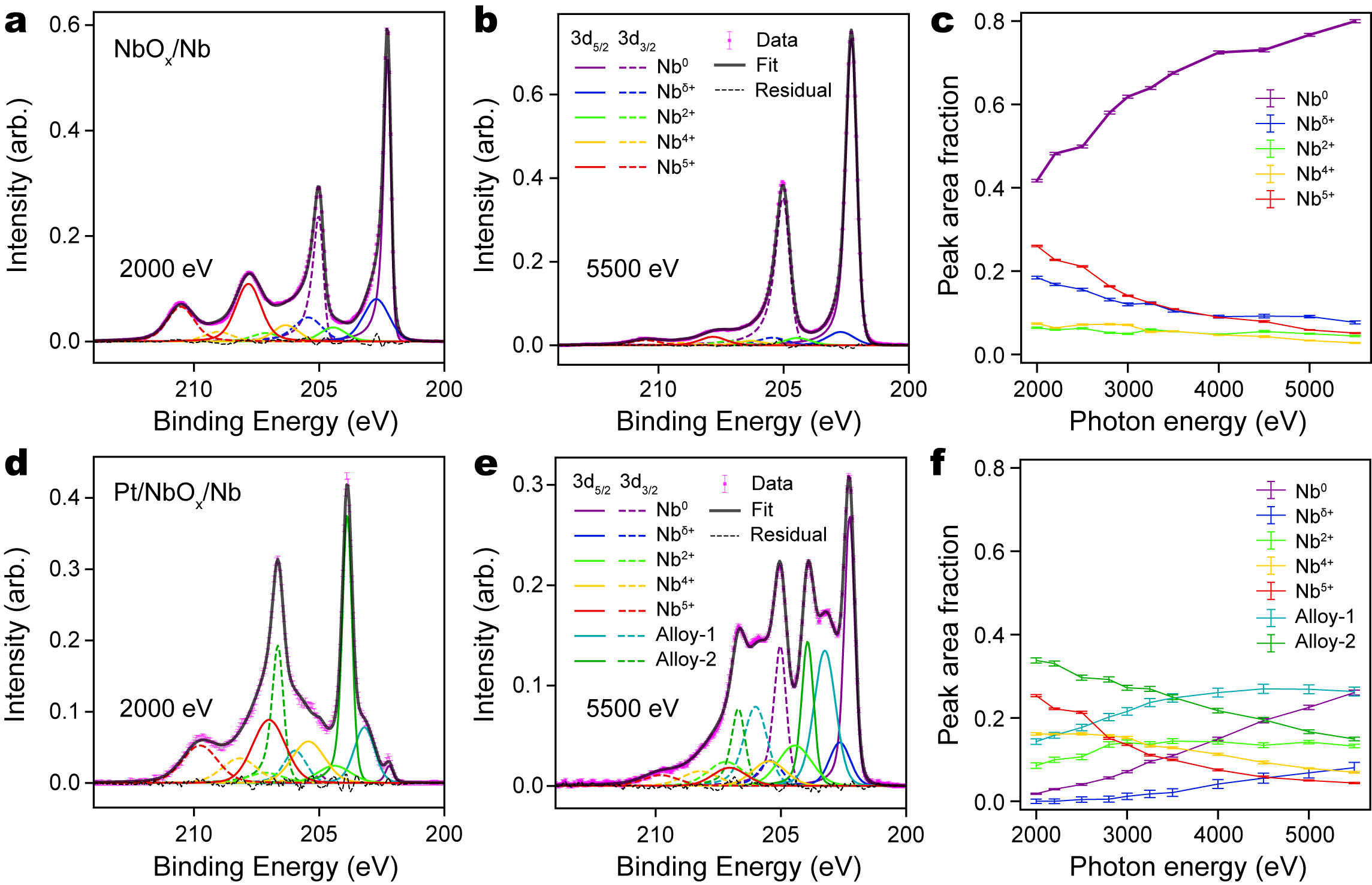}

\caption{(a, b) Nb 3d core-level VEXPS spectra of bare Nb measured at photon energies of 2000 eV and 5500 eV, respectively, together with the best peak fitting results and residuals. (c) The Nb 3d peak area fractions as functions of photon energy for bare Nb film. (d,e) Nb 3d core-level VEXPS spectra of Pt/NbO$_x$/Nb measured at 2000 eV and 5500 eV, respectively, together with the best peak fitting results and residuals. (f) The Nb 3d peak area fractions as functions of photon energy for Pt/NbO$_x$/Nb film. }
\label{fig3:HAXPES}
\end{figure}

To further elucidate the chemical states associated with the surface region, XPS analyses of both NbO$_x$/Nb and Pt/NbO$_x$/Nb films were first carried out using a laboratory-scale X-ray source (Al K-$\alpha$) and subsequently at a synchrotron facility. The surface chemistry of Nb is known to be complex due to the coexistence of multiple native oxide phases, which is further complicated by the formation of Nb–Pt alloys. Overlapping spectroscopic signatures of multiple co-existing Nb oxidation states render conventional single-spectrum XPS peak deconvolution unreliable, as it inevitably leads to overfitting. To obtain a robust and physically justified spectral deconvolution, we employ variable energy X-ray photoelectron spectroscopy (VEXPS) using 11 photon energies between 2000 eV and 5500 eV (Figures \ref{fig3:HAXPES} and S3,4). Increasing the photon energy increases the photoelectron kinetic energy and consequently the effective probing depth, making the technique progressively more sensitive to bulk species. Accordingly, the measured spectra can be described as a linear combination of chemically distinct components with photon-energy-dependent weights. By fitting the entire VEXPS spectral series using an identical set of spectral components while allowing their relative weights to vary with photon energy, we achieve substantially higher confidence in identifying chemical species near the surface and quantifying their concentrations. Details of the VEXPS fitting methodology, including line-shape selection and parameter constraints, are provided in the Supporting Information.

For the \ce{NbO_x}/Nb film, two major components can be readily identified from their characteristic Nb 3d$_{5/2}$–3d$_{3/2}$ doublets: metallic Nb (\ce{Nb^0}, binding energies approximately 202.2 eV and 204.9 eV) and \ce{Nb2O5} (\ce{Nb^{5+}}, 207.8 eV and 210.5 eV), together with additional shoulder features attributable to Nb suboxides and interfacial species (Figures \ref{fig3:HAXPES}a, b) \cite{Premkumar2021Microscopic, Kar2023NeutralAtomBeam, Prudnikava2022NbThermalTreatments, dacca1998xps,nwanna2020nbxoy}. According to the global VEXPS deconvolution, these shoulder features can be fully accounted for by introducing three additional chemically distinct Nb species. The three species exhibit Nb 3d binding energy shifts of +4.17 eV, +2.32 eV, and +0.52 eV relative to \ce{Nb^0}. By comparison with literature values, the first two components are readily identified as \ce{Nb^{4+}} (\ce{NbO2}) and \ce{Nb^{2+}} (\ce{NbO}), the two stable suboxides of Nb \cite{dacca1998xps,nwanna2020nbxoy}. The third species, exhibiting only a small chemical shift, is therefore unlikely to correspond to a stoichiometric Nb oxide, but instead reflects a solid solution of oxygen in metallic Nb, as permitted by the Nb–O phase diagram \cite{naito1984nbO}. In this case, oxygen atoms occupy interstitial sites within the bcc Nb lattice, locally modifying the electronic environment of Nb while preserving its metallic character, most plausibly near the metal–oxide interface. This species is labeled as \ce{Nb^{\delta+}} in the following discussion. The binding energies of all oxidized Nb species and other fitted parameters are summarized in Supporting Information, Tables SI and SII, and are found to be in good agreement with reported literature values.

To obtain a deeper insight into the relationship between the local atomic structure and Nb oxidation states, we turn to ab initio simulations. The partially oxidized Nb(111) surface was modeled by adsorbing oxygen atoms on the surface and incorporating them into the subsurface region; followed by the static total energy minimization with respect to the atomic coordinates. The cutoff radius of 2.8 \AA\ (Figure S5) was used to determine the number of oxygen neighbors near each Nb atom in the resulting NbO$_x$/Nb structure. The Nb atomic charges vary between approximately +0.1 for the Nb atoms in the metallic part of the NbO$_x$/Nb slab and approximately +2.4 for 5-coordinated Nb species. The small positive charge on the Nb atoms that have no oxygen atoms in their vicinity is consistent with the Nb$^{\delta+}$ species resolved experimentally; it indicates that the oxygen impurities affect the charge state of the host lattice atoms at the distance of at least 1 nm. The general trend of a higher Nb oxidation state with the larger number of oxygen atoms is evident (Figure S5), however, the values of charge can vary within $\pm$0.5 depending on the specific arrangements of the Nb and O atoms. 

The Nb 3d XPS of the Pt/NbO$_x$/Nb film is considerably more complex than that of NbO$_x$/Nb due to the formation of Nb–Pt alloys (Figures 3d,e), which give rise to additional spectral features, including sharp peaks at approximately 203.9 eV and 206.6 eV. These features, attributed to the alloy formation, partially overlap with the oxide peaks, such that only one oxide-related peak, presumably associated with the 3d$_{3/2}$ branch of an Nb 3d doublet, can be visually resolved near 210.0 eV. Further insight into the Nb 3d spectra of Pt/NbO$_x$/Nb is obtained through global deconvolution of the VEXPS spectral series. To achieve satisfactory fits, spectral contributions due to all chemical species identified for the NbO$_x$/Nb film are retained, and two additional components are introduced to account for Nb–Pt alloy formation. These two alloy components exhibit Nb 3d binding energy shifts of +0.95 eV (Alloy-1) and +1.67 eV (Alloy-2) relative to metallic Nb. The positive chemical shifts are attributed to the substantially higher electronegativity of Pt compared to Nb (2.28 vs.\ 1.6 on the Pauling scale), which drives charge transfer from Nb to Pt and results in an increased Nb 3d binding energy. Notably, the peaks associated with \ce{Nb^{5+}} and \ce{Nb^{4+}} in the Pt/NbO$_x$/Nb film are shifted to lower binding energies compared to those in the NbO$_x$/Nb film. For NbO$_x$/Nb, the binding energies of \ce{Nb^{5+}} and \ce{Nb^{4+}} are shifted by +5.62 eV and +4.12 eV from \ce{Nb^0}, respectively, whereas in the Pt/NbO$_x$/Nb film these shifts decrease to +4.83 eV and +3.26 eV. This red-shift of approximately 0.8 eV indicates that \ce{Nb2O5} and \ce{NbO2} are partially reduced in the Pt/NbO$_x$/Nb thin film, suggesting oxygen diffusion during the formation of Nb-Pt alloys.

The photon-energy dependence of VEXPS peak fractions provides a powerful means of assessing the spatial distribution of chemical species within the thin-film stack. Because photoelectrons with higher kinetic energies have longer inelastic mean free paths, the spectral weight of surface species decreases with photon energy, whereas buried species exhibit increasing relative intensity. For the \ce{NbO_x}/Nb film, all oxidized components (\ce{Nb^{5+}}, \ce{Nb^{4+}}, \ce{Nb^{2+}}, and \ce{Nb^{\delta+}}) behave as surface species, while only the bulk-dominant \ce{Nb^0} behaves as buried, as expected (Figure \ref{fig3:HAXPES}c). Applying the same analysis to the Pt/NbO$_x$/Nb film reveals distinctly different depth profiles for the two Nb–Pt alloy components (Figure \ref{fig3:HAXPES}f): Alloy-1 appears buried, whereas Alloy-2 appears as surface-dominant. Among the oxidized species, the partially reduced \ce{Nb^{5+}} and \ce{Nb^{4+}} components remain surface-dominant, whereas \ce{Nb^{\delta+}} appears buried. In contrast, the spectral weight of \ce{Nb^{2+}} initially increases with photon energy but decreases again above 4000 eV, indicating that this species is preferentially located at an intermediate depth, likely near the interface between the bulk metal and the surface oxide. The concurrent inward migration of \ce{Nb^{\delta+}} and \ce{Nb^{2+}} is consistent with the diffusion of oxygen that accompanies the partial reduction of \ce{Nb2O5} and \ce{NbO2}. 

The static energy minimization of an atomic-scale model of Pt coating on the \ce{NbO_x}/Nb films reveals mechanistic features consistent with these experimental observations (Figure 4 and Figure S6). The Pt layer smears out along the NbO$_x$ surface and forms a stack of deformed Pt(111) planes, indicating a strong interaction with the partially oxidized Nb. The energy gain associated with the attachment of these Pt planes to the \ce{NbO_x} surface is approximately 4.5 J/m$^2$, i.e., significantly larger than binding energies of metal clusters to oxide substrates~\cite{hemmingson2017trends}. This attachment energy increases to 5.4 J/m$^2$ if oxygen atoms are allowed to spread out over the Nb region (Figure 4a and Figure S6).

According to the population analysis, Pt deposition induces accumulation of additional positive charge on the Nb atoms, which trends higher for more dispersed oxygen distributions (Figure 4b and Figure S6), but remains the same for 1 and 3 ML (monolayers) Pt coverage. Oxygen spread in Figure 4 is defined as the difference between the $max$ and $min$ oxygen locations over the Nb region (Supporting Information, Figure S6). In contrast, the total oxygen atomic charge shows a strong dependence on the oxygen spread but appears insensitive to the Pt coverage (Figure 4c). This distribution suggests electron transfer from Nb to the Pt layer. Indeed, our analysis suggests that Pt atoms in the immediate vicinity of the NbO$_x$ substrate acquire the charge between $-0.1\,e$ and $-0.7\,e$ depending on the oxygen distribution, while atoms in the second and third to the interface planes remain neutral (Figure 4e and Figure S7). 

The Nb-to-Pt electron transfer creates local electric fields at the Pt/\ce{NbO_x} and \ce{NbO_x}/Nb interfaces (Figures 4e, insert) that promotes atomic redistribution. In particular, negatively charged oxide ions displace away from the Pt layer deeper into the Nb bulk. As a result, displaced oxygen species oxidize Nb in a wider near-interface region, while Nb atoms in the outermost region of the NbO$_x$ film become reduced. We note that in the absence of Pt (0 ML in Figure 4d), such dispersion of the oxide layer is thermodynamically unfavorable, consistent with the absence of the internal electrostatic field. However, in the presence of Pt, it becomes cost neutral regardless of the Pt layer thickness.

In addition to the oxide ion diffusion, the internal electric field promotes the displacement of Nb cations toward negatively charged Pt, thereby facilitating further dilution of the oxygen content near the Pt/\ce{NbO_x} interface. Further stabilization of this interface stems from alloying of the Pt coating with interfacial Nb that was depleted of oxygen. According to our calculations, energy gain due to alloying Pt into the bcc lattice of Nb (Figure 4f and, Figure S8) reaches up to 0.4 eV per atom. 
The calculated charge distribution in these model Nb-Pt alloys is consistent with their relative electronegativities (Figure S7). While Nb ionic charges vary between $0\,e$ and $1.5\,e$ depending on the local Pt content, we can identify charge states corresponding to characteristic cases: Nb surrounded by Pt (+1.5), Nb at a Pt/Nb interface (+1), and Nb in the presence of dilute concentration of Pt ($<$ +0.5) (Figure S7).

For comparison, Nb-Pt alloying energy calculated for the crystalline NbPt$_x$ phases (Figure 4g) follows the same profile but reaches the value --0.7 eV. This comparison suggest that approximately half of the Nb-Pt intermixing energy stems from the charge redistribution, while the other half is associated with reorganization of the local structure. While this is a significant contribution, since the NbPt, NbPt$_2$, and NbPt$_3$ have similar alloying energies (Figure 4g), it can be expected that the Nb-Pt alloyed region is either amorphous or contains a combination of these intermetallic compounds.

 Given that the HAADF-STEM analysis reveals a higher Pt concentration near the film surface (Figure \ref{fig4:XAFS}a), Alloy-2 can therefore be assigned to a more Pt-rich alloy environment than Alloy-1, consistent with its larger Nb 3d chemical shift (+1.67 eV for Alloy-2 versus +0.95 eV for Alloy-1). While STEM–EDX and VEXPS analysis both identify alloy formation and oxygen diffusion after Pt encapsulation, they do not directly determine the atomic coordination environment of Pt within the reacted layer. To resolve the dominant local structure of the surface layer, extended X-ray absorption fine structure (EXAFS) spectra are collected from the Pt/NbO$_x$/Nb film at grazing incidence, for enhanced surface sensitivity. As expected, the Nb $K$-edge EXAFS spectrum is well described by a bulk bcc structural model, due to the technique's bulk sensitivity. However, the Pt $L_3$-edge EXAFS is clearly different from bulk fcc Pt. The $k^3$-weighted Pt $L_3$-edge EXAFS oscillations extend to $k\approx 14.5$ \AA$^{-1}$, indicating a well-defined and highly ordered local coordination environment rather than a disordered reaction layer. When the spectrum is fitted using an fcc Pt structural model, catastrophic disagreement is obtained, with a reduced $\chi^2$ approaching 100 and an R--factor of 0.27 (Figure S9). The result provides direct spectroscopic evidence of substantial Nb--Pt alloying, so that the local environment of Pt in the Pt/NbO$_x$/Nb film becomes fundamentally different from bulk metallic Pt. 

Quantitative EXAFS spectral fitting was performed using crystallographic Nb--Pt alloy models including \ce{NbPt3} (mp-12700), \ce{NbPt2} (mp-11514), and NbPt (mp-999376) \cite{Jain2013}. Among these candidates, the Pt-rich \ce{NbPt3} model reproduces both the $k$--space oscillations and the $R$--space Fourier transform with high fidelity (Figures \ref{fig4:XAFS}b,c). The fitting yields a reduced $\chi^2$ of 2.38 and an exceptionally low $R$--factor of $5.6\times10^{-3}$ while maintaining physically realistic structural parameters. The refined first-shell bond lengths are $(2.756 \pm 0.001)$ \AA\ for Pt--Pt and $(2.781 \pm 0.003)$ \AA\ for Pt--Nb, closely resembling the bond lengths in crystalline \ce{NbPt3}. Importantly, the number of independent data points substantially exceeds the number of refined variables, ensuring that the fit is statistically robust and not over-parameterized. In contrast, the \ce{NbPt2} model produces elevated residuals and fails to reproduce higher-shell features (Figure S10), suggesting that it at most represents a minor secondary environment. The NbPt model, on the other hand, yields unphysical structural parameters (Figure S11), including an unrealistically short Pt--Nb bond length near 1.99 \AA\ and unstable Debye--Waller factors, rendering this configuration chemically implausible. These systematic comparisons conclusively establish that the surface Nb--Pt alloy is dominated by \ce{NbPt3}, corresponding to the Pt-rich, surface-dominant Alloy-2, while \ce{NbPt2} likely contributes to the buried Alloy-1. The Nb--Pt alloy energy versus composition is shown in Figure 4. Note that \ce{NbPt2} is the most stable; therefore, its formation deeper in the film, where the availability of Nb is higher, suppresses Nb migration toward the surface, leaving the surface Pt-rich (\ce{NbPt3}). Fitting of the Nb K-edge EXAFS was also carried out and is presented in Supporting Information, Figure S12, indicating that the bulk characteristics of Nb remain unchanged, with no significant variations observed.

The mechanisms of transformations corresponding to the Nb surface processing discussed here are summarized in Figure 6. Exposure of as deposited Nb surface to ambient environment results in the formation of native oxide NbO$_x$. Subsequent deposition of Pt induces the electron transfer from subsurface Nb to the interfacial Pt plane. The resulting electric field promotes the diffusion of positively charge Nb ions toward the Pt layer, while the O$^{2-}$ diffuse deeper into the Nb film. While the oxygen dissolution into the Nb is nearly thermodynamically  cost-neutral, oxygen depletion near the Pt/NbO$_x$ interface allows for excess Nb to intermix with Pt with the energy gain of up to 0.7 eV per atom. These oxygen dissolution and Nb-Pt alloying processes can proceed until the Nb-Pt layer reaches the most stable average composition (NbPt$_2$ -- NbPt$_3$), with the widths of both Nb-Pt alloy and Nb-O solid solution layers determined by the annealing temperature and treatment duration.

Taken together, these results reveal a vertically graded, Pt-rich \ce{NbPt3} interfacial alloy formed during annealing that simultaneously reduces the native Nb oxide while preserving the bulk superconducting properties of Nb. The process is therefore not merely protective, but chemically transformative, converting an oxide-covered surface into a metallically bonded alloy interface. Because native Nb oxide layer is implicated in dielectric loss in superconducting circuits, chemically transforming this layer into an inert metallic region may offer a promising strategy for improving coherence times. In addition, the relatively long duration (1 -- 2 hours) of annealing at $750^\circ$C represents a significant thermal budget for fabrication scaling up. In future, techniques that have lower thermal budget, such as flash lamp annealing and pulsed laser annealing, may serve as more desirable alternatives.

\section*{Conclusions}

In summary, we demonstrate a novel encapsulation strategy for transforming the native oxide on superconducting Nb into a chemically stable, metallic alloy-dominated surface. Structural and chemical characterization, together with density functional theory calculations, establishes the atomic-scale evolution of the interface and the thermodynamic driving force for the Pt--Nb alloy formation and the inward oxygen diffusion, while DC transport measurements confirm that the bulk superconducting properties of the underlying Nb are preserved. More broadly, this work establishes post-oxidation interfacial reconstruction as a materials strategy for converting native oxide-covered Nb into a chemically robust alloyed surface, complementing conventional approaches that rely on preventing oxidation of pristine films. Looking forward, given that sputter deposition is less directional, the approach may offer a route toward protecting etched Nb sidewalls as well, although substantial process optimization and validation will be needed. Determining the microwave-loss characteristics of the reconstructed interface, including the potential role of redistributed interstitial oxygen, will be an important subject of future device-level studies.

\section*{Methods}

\subsection{Materials fabrication.}
Nb thin films were deposited in a magnetron sputtering system (AJA International) that was pumped to a base pressure below $2.7\times10^{-7}$ Pa. Single-side polished c-cut sapphire (\ce{Al2O3}) substrates (CrysTec) were used for film growth. Prior to deposition, the substrates were sequentially cleaned using a Piranha solution (\ce{H2SO4}:\ce{H2O2} = 2:1), followed by rinsing with isopropyl alcohol and deionized water. During deposition, the substrate temperature was maintained at 700 $^\circ$C using a radiatively heated stage. After the deposition of Nb, the substrate holder was cooled to room temperature, removed from the deposition chamber, and exposed to ambient atmosphere for 1 hour. After the formation of native oxide, the Nb-covered wafer was loaded back into the deposition chamber. Subsequently, a Pt capping layer was sputter deposited at room temperature, followed by an \emph{in situ} vacuum annealing at 750 $^\circ$C for 2 hours. The sample was then cooled to room temperature and removed from the deposition chamber. Magnetron sputtering of Nb and Pt was carried out at powers of 150 W and 100 W, respectively to thicknesses of about 110 nm and 10 nm, under 1.6 Pa of Ar. Nb and Pt sputtering targets (99.995\% purity) were sourced from AJA International.

\subsection{Materials Characterizations}

Low-temperature electrical transport was measured in a Quantum Design Dynacool PPMS system, using its built-in DC resistivity option. The superconducting transition temperature (Tc) was defined as the midpoint of the resistive transition ($R/R_n = 0.5$, with $R_n$ measured at 10 K). The residual resistivity ratio (RRR) was determined as the ratio of the room-temperature resistance to $R_n$. To assess reproducibility, three independent samples were characterized under each processing condition. The reported uncertainties were calculated according to the measurement type. For the superconducting critical temperature, the uncertainty was estimated from the temperature step size used during the measurement. Since the temperature step size was 0.1 K, the uncertainty in $T_c$ was conservatively taken as one temperature step, and the transition temperature is reported as $T_c = (9.30 \pm 0.10$) K. This uncertainty represents the experimental temperature resolution of the measurement.

X-ray diffraction (XRD) patterns of the thin film samples were collected with a Rigaku SmartLab diffractometer equipped with a Cu $K_\alpha$ radiation source ($\lambda = 1.5418$ \AA). Grazing-incidence XRD (GIXRD) measurements were conducted at a fixed incident angle of $0.5^\circ$ to enhance surface sensitivity. 

Transmission electron microscopy (TEM) lamellae were prepared using a Thermo Fisher Scientific Helios G5 dual-beam focused ion beam (FIB) system employing an in situ lift-out procedure. Final thinning to  approximately 50 nm was achieved through progressively lower \ce{Ga^+} beam voltages to minimize ion-beam damage. TEM imaging and analysis were performed on a Thermo Fisher Scientific Talos X-FEG 200X microscope operated at 200 kV and equipped with a Super-X EDS system. High-angle annular dark-field scanning transmission electron microscopy (HAADF-STEM) and electron energy-loss spectroscopy (EELS) were carried out on a Hitachi HD2700C microscope with probe Cs correction and a Gatan Enfinium spectrometer operated at 200 kV. Film thicknesses, interplanar spacings, and mapping regions were quantified using ImageJ.

Variable-energy X-ray photoelectron spectroscopy (VEXPS) measurements were carried out at the hard X-ray photoelectron spectroscopy (HAXPES) endstation of the Spectroscopy Soft and Tender (SST) beamline suite, 7-ID, at the National Synchrotron Light Source II (NSLS-II), Brookhaven National Laboratory, using photon energies in the tender X-ray range from 2.0 keV to 6.0 keV from the SST-2 source. A double-crystal monochromator with Si(111) crystals was used for photon energies between 2.0 keV and 3.0 keV, and Si(220) crystals were used for photon energies above 3.25 keV. The photoelectron take-off angle was fixed at $80^\circ$, with a $10^\circ$ X-ray incidence angle. Energy calibration was performed using a gold thin film in electrical contact with the sample holder by setting the Au $4f_{7/2}$ binding energy to 84.0 eV, and background subtraction was carried out using a Shirley model.

Extended X-ray absorption fine structure (EXAFS) measurements were performed at the Beamline for Materials Measurement (BMM), 6-BM, at the National Synchrotron Light Source II (NSLS-II), Brookhaven National Laboratory. The incident beam was conditioned by a collimating mirror, a Si(111) double-crystal monochromator, and a downstream harmonic-rejection mirror. Measurements were carried out in fluorescence mode using a grazing-incidence geometry ($3^\circ$ incidence) to enhance surface sensitivity, and the samples were continuously rotated about their surface normal to minimize Bragg diffraction. Spectra were collected at the Nb $K$-edge and Pt $L_3$-edge, and multiple scans were acquired and averaged to improve the signal-to-noise ratio. Nb and Pt foils were used for energy calibration. X-ray absorption spectra were processed in Athena, and EXAFS fitting was carried out in Artemis within the Demeter package. \cite{Ravel2005Athena} 

\subsection{Ab Initio Modeling}

The pure, oxidized, and Pt-coated Nb(111) surfaces were represented using the periodic slab model. The Nb slab was 12 atomic planes thick, with four non-equivalent Nb atoms per plane. The in-plane supercell parameters are $a=b=9.33$~\AA, $\gamma=120^{\circ}$ in crystallographic notations. The out-of-plane supercell parameter $c=30$~\AA, leaving an over 13~\AA\ vacuum gap after the oxidation-induced expansion and Pt coating are taken into account. Nb atoms of the bottom plane of the slab were fixed in their bulk atomic positions. Coordinates of all of the other atoms were fully relaxed. The calculations were performed using the VASP code \cite{kresse1996vasp} 
 and Perdew-Burke-Ernzerhof \cite{pbe1996} exchange correlation functional. The projector-augmented wave potentials were used to approximate the effect of the core electrons \cite{blochl1994paw}. The calculations were carried out using the $\Gamma$-centered $3\times3\times1$ $k$-mesh and the plane-wave basis set cutoff of 500 eV. The total energy convergence criterion was set to $10^{-5}$~eV. Bader charge population analysis was used to analyze the charge density redistribution \cite{henkelman2006bader}. 

\begin{acknowledgments}

This material is based upon work supported by the U.S. Department of Energy, Office of Science, National Quantum Information Science Research Centers, Co-design Center for Quantum Advantage (C2QA) under contract number DE-SC0012704 and PNNL FWP 76274. This research used resources of the National Synchrotron Light Source II (NSLS-II), including the Submicron Resolution X-ray Spectroscopy beamline (SRX), the Hard X-ray Nanoprobe beamline (HXN), the National Institute of Standards and Technology (NIST)-operated Spectroscopy Soft and Tender beamlines (SST-1/SST-2) , and the EXAFS. International Business Machines Corporation (IBM) operates the diffractometer. This work also used the Materials Synthesis \& Characterization, Proximal Probes, Nanofabrication, and Electron Microscopy facilities of the Center for Functional Nanomaterials (CFN). NSLS-II and CFN are both U.S. Department of Energy Office of Science User Facilities at Brookhaven National Laboratory, operated under Contract No.\ DE-SC0012704. This research used resources of the National Energy Research Scientific Computing Center, a DOE Office of Science User Facility supported by the Office of Science of the U.S. Department of Energy under Contract No. DE-AC02-05CH11231 using NERSC award BES-ERCAP0037279. 

\end{acknowledgments}

\section*{Disclaimer}

Certain commercial equipment, instruments, software, or materials are identified in this paper in order to specify the experimental procedure adequately, and do not represent an endorsement by the National Institute of Standards and Technology.

\section*{Author Declaration}

Conflict of Interest: The authors have no conflicts to disclose.

\section*{Author Contribution}

A.C. and M.L. (Mingzhao Liu) conceived the project. A.C. prepared the thin film samples. M.L. (Ming Lu) and N.T. supported the sample fabrication. A.C., C.W., A.P., A.M.K., and A.K.A. performed the hard X-ray photoelectron spectroscopy (HAXPES) measurements. A.C., A.K.A, and B.R. performed the extended X-ray absorption fine structure (EXAFS) measurements. A.C. and X.T. carried out the laboratory-based X-ray photoelectron spectroscopy (XPS) measurements. K.K. and S.H. performed the electron microscopy characterization. P.V.S. carried out the first-principles calculations. A.C., A.K.A., A.L.W., and M.L. (Mingzhao Liu) analyzed and interpreted the experimental data. A.C., A.K.A., A.L.W., P.V.S., and M.L. (Mingzhao Liu) wrote the manuscript with input from all authors. S.L.H. and M.L. supervised the project and acquired funding. All authors discussed the results and approved the final manuscript.

\section*{Data Availability Statement}

Data generated and analyzed under this study is included in this article and available on reasonable request.

\newpage

\newpage

\newpage

\newpage

\begin{figure}[p]
\centering
\includegraphics[width=0.9\columnwidth]{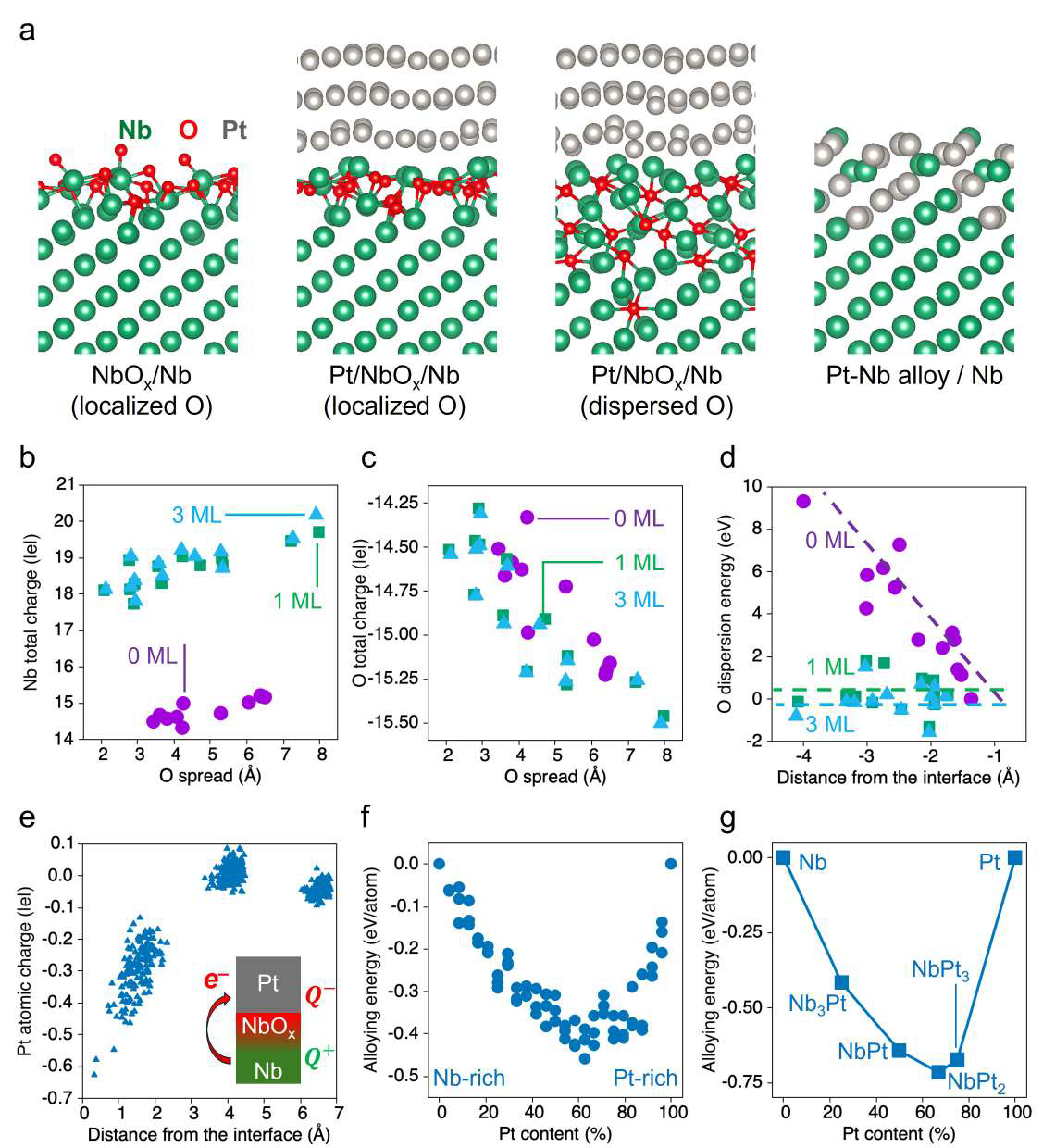}
\caption{
(a) Computational models of representative structures. From left to right: partially oxidized Nb (NbO$_x$/Nb); Pt (3 ML)/NbO$_x$/Nb with localized and dispersed distributions of oxygen near the Pt/NbO$_x$ interface; and an Nb-Pt alloy configuration (see also Figure SI 7). 
(b,c) Total charges of Nb and O sub-lattices versus the spread of the oxygen atoms in the NbO$_x$ layer without (0 ML) and with (1 ML, 3 ML) Pt film.  
(d) Energy cost associated dispersion of oxygen species over NbO$_x$ without Pt (0 ML) and with Pt (1 ML, 3 ML) films. See Figure SI 6 for the discussion of dispersed O configurations.
(e) Pt atomic charges for 3 ML Pt film. Inset illustrates Nb-to-Pt charge transfer occurs to the interfacial Pt plane only.
(f) Alloying energy due to Nb-Pt intermixing in the bcc Nb slab. See Figure SI 7.
(g) Alloying energy due to the formation of crystalline phases of NbPt$_x$ intermetallic compounds.}
\label{fig5:alloy_cartoon}
\end{figure}

\begin{figure}
\includegraphics[width=1.0\columnwidth]{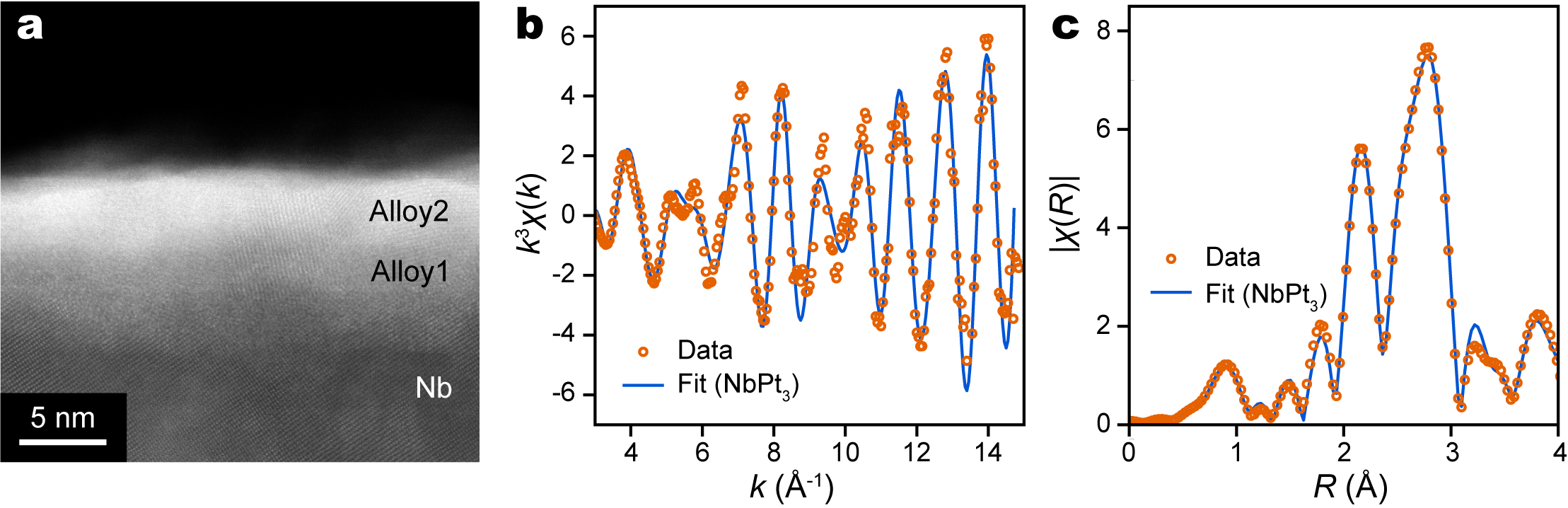}
\caption{(a) Cross-sectional HAADF-STEM image of the top surface of Pt/NbO$_x$/Nb thin film. The $Z$-contrast transition along the film depth reveals two intermetallic alloy layers stacking on top of bulk Nb. By comparison with the HAXPES fitting results, the top-most layer is identified as Nb-Pt Alloy-2, while the layer below is identified as Nb-Pt Alloy-1. (b) $k^3$-weighted Pt $L_3$-edge EXAFS oscillations, $\chi(k)\cdot k^3$, measured in a $3^\circ$ grazing-incidence geometry (circles), overlapped with the best fit using the \ce{NbPt3} structure (solid line). (c) Magnitude of the Fourier-transformed $R$-space EXAFS signal $|\chi(R)|$ with the fit performed over 1.6~\AA{} to 4.1~\AA{} using the same \ce{NbPt3} structure. The refined first-shell Pt--Pt and Pt--Nb bond lengths are $(2.756 \pm 0.001)$ \AA\ and $(2.781 \pm 0.003)$ \AA, respectively.}
\label{fig4:XAFS}
\end{figure}

\begin{figure}
\includegraphics[width=1.0\columnwidth]{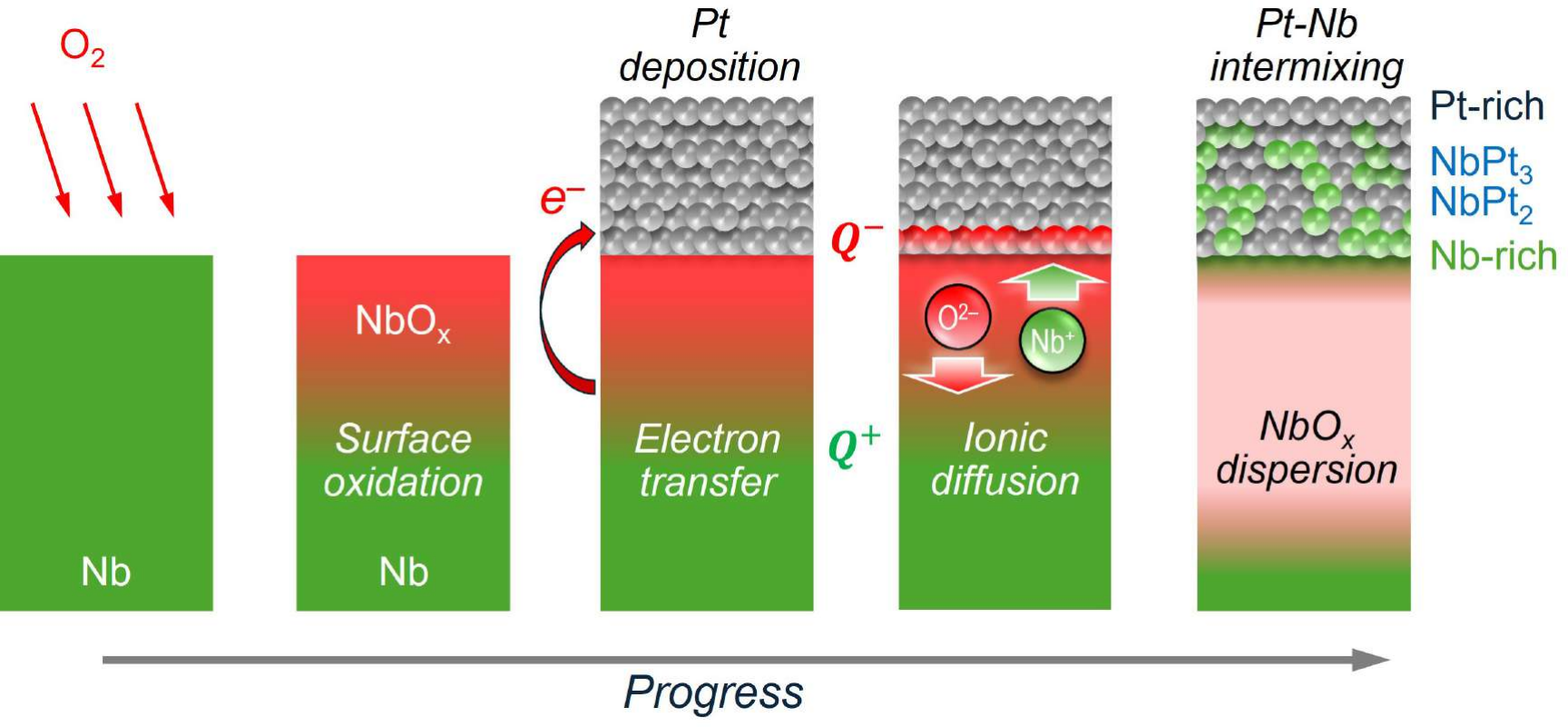}
\caption{The mechanism of Nb-Pt alloy formation and oxygen dissolution in the Nb film: Native surface oxidation initially forms an NbO$_x$ layer on Nb upon O$_2$ exposure. During Pt deposition, charge transfer occurs across the Pt/NbO$_x$ interface, followed by ionic diffusion and Nb-Pt intermixing during annealing. The resulting interfacial structure evolves into Nb-Pt alloying compounds, while oxygen remains dissolved within the Nb matrix rather than forming a separate oxide phase. This encapsulation-driven interfacial evolution suppresses further oxidation and stabilizes the superconducting Nb surface for qubit-device applications.}

\end{figure}

\clearpage
\onecolumngrid
\appendix
\section*{Supporting Information}

\title{Supporting Information for ``Transforming Native Oxide into a Metallic Platinum--Niobium Alloy for Passivation of Superconducting Niobium Films''}

\author{Ananya Chattaraj\,\orcidlink{0000-0002-9225-5224}}
\affiliation{Center for Functional Nanomaterials, Brookhaven National Laboratory, Upton, NY 11973, USA.}%

\author{Conan Weiland\,\orcidlink{0000-0001-6808-1941}}
\affiliation{Material Measurement Laboratory, National Institute of Standards and Technology, Gaithersburg, MD, 20899 USA}

\author{Bruce Ravel\,\orcidlink{0000-0002-4126-872X}}
\affiliation{Material Measurement Laboratory, National Institute of Standards and Technology, Gaithersburg, MD, 20899 USA}

\author{Kim Kisslinger\,\orcidlink{0000-0002-6528-7044}}
\affiliation{Center for Functional Nanomaterials, Brookhaven National Laboratory, Upton, NY 11973, USA.}%

\author{Sooyeon Hwang\,\orcidlink{0000-0001-5606-6728}}
\affiliation{Center for Functional Nanomaterials, Brookhaven National Laboratory, Upton, NY 11973, USA.}%

\author{Ming Lu\,\orcidlink{0000-0003-0175-6531}}
\affiliation{Center for Functional Nanomaterials, Brookhaven National Laboratory, Upton, NY 11973, USA.}%

\author{Nikhil Tiwale\,\orcidlink{0000-0001-8229-7108}}
\affiliation{Center for Functional Nanomaterials, Brookhaven National Laboratory, Upton, NY 11973, USA.}%

\author{Xiao Tong\,\orcidlink{0000-0002-5567-9677}}
\affiliation{Center for Functional Nanomaterials, Brookhaven National Laboratory, Upton, NY 11973, USA.}%

\author{Ajith Pattammattel\,\orcidlink{0000-0002-5956-7808}}
\affiliation{National Synchrotron Light Source II, Brookhaven National Laboratory, Upton, NY 11973, USA.}%

\author{Andrew M. Kiss\,\orcidlink{0000-0002-8515-5508}}
\affiliation{National Synchrotron Light Source II, Brookhaven National Laboratory, Upton, NY 11973, USA.}%

\author{Steven L. Hulbert\,\orcidlink{0000-0003-3127-6029}}
\affiliation{National Synchrotron Light Source II, Brookhaven National Laboratory, Upton, NY 11973, USA.}%

\author{Aswin kumar Anbalagan\,\orcidlink{0000-0001-5511-2083}}
\email{Contact author: aanbalaga1@bnl.gov}
\affiliation{National Synchrotron Light Source II, Brookhaven National Laboratory, Upton, NY 11973, USA.}%

\author{Andrew L. Walter\,\orcidlink{0009-0002-4215-8546}}
\email{Contact author: awalter@bnl.gov}
\affiliation{National Synchrotron Light Source II, Brookhaven National Laboratory, Upton, NY 11973, USA.}%

\author{Peter V. Sushko\,\orcidlink{0000-0001-7338-4146}}
\email{Contact author: peter.sushko@pnnl.gov}
\affiliation{Integrated Discovery Sciences Directorate, Pacific Northwest National Laboratory, Richland, WA 99354, USA.}%

\author{Mingzhao Liu\,\orcidlink{0000-0002-0999-5214}}
 \email{Contact author: mzliu@bnl.gov}
\affiliation{Center for Functional Nanomaterials, Brookhaven National Laboratory, Upton, NY 11973, USA.}%
\date{\today}

\maketitle


\renewcommand{\thetable}{S\arabic{table}}
\setcounter{table}{0}
\setcounter{figure}{0}
\renewcommand{\thefigure}{S\arabic{figure}}

\section{X-ray Diffraction (XRD) Study and Analysis}

\begin{figure}[htbp]
\includegraphics[width=0.9\columnwidth]{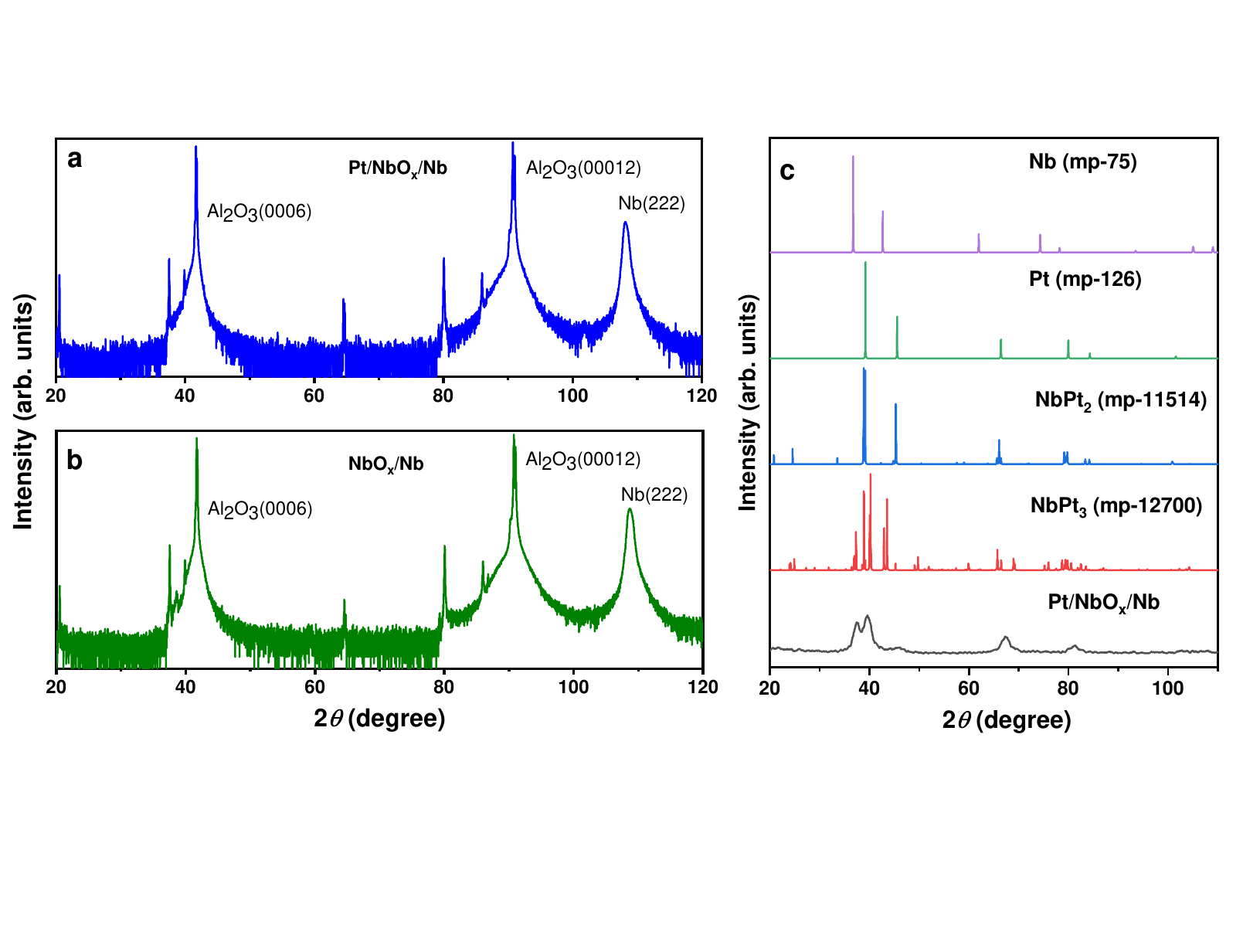}
\caption{(a) $\theta$--2$\theta$ X-ray diffraction (XRD) pattern of Pt/NbO$_x$/Nb thin films on Al$_2$O$_3$. 
(b) $\theta$--2$\theta$ XRD pattern of bare NbO$_x$/Nb on Al$_2$O$_3$ measured under identical conditions. 
Both patterns are dominated by Nb and substrate reflections, showing minimal differences due to the limited thickness of the Pt overlayer and interfacial region. 
(c) Grazing-incidence XRD (GIXRD, incidence angle = $1^\circ$) of Pt/NbO$_x$/Nb compared with reference patterns of Nb, Pt, NbPt$_2$, and NbPt$_3$ from the Materials Project database.
The GIXRD pattern exhibits weak and broadened features that do not fully match bulk reference patterns but are consistent with Nb--Pt alloy formation.}
\label{FIG_SI_1}
\end{figure}

The $\theta$--2$\theta$ XRD patterns of Pt/NbO$_x$/Nb and NbO$_x$/Nb are nearly identical, indicating that the bulk crystal structure of Nb remains largely unchanged upon Pt encapsulation. This is expected given the limited thickness of the Pt overlayer and interfacial region relative to the Nb film. In contrast, Grazing-Incidence X-ray Diffraction (GIXRD) measurements, which are more surface-sensitive, reveal weak and broadened features that do not match pure Pt or Nb phases but show partial correspondence with NbPt$_2$- and NbPt$_3$-related reflections. These observations suggest the formation of a thin, structurally disordered or nanocrystalline Nb--Pt alloy region at the surface, rather than well-crystallized bulk intermetallic phases.

\clearpage

\section{Thickness Calculations from Transmission Electron Microscopy (TEM)}

TEM measurements show that the Nb film without Pt capping has an average thickness of (113.8 $\pm$ 0.9) nm. In the Pt-capped sample, the measured Nb thickness of (107.9 $\pm$ 1.7) nm reflects only the remaining chemically distinct Nb layer, while Nb incorporated into the Pt--Nb interfacial alloy region is not counted. Thus, the lower apparent thickness does not indicate a reduction in total Nb content. The native Nb oxide layer is (2.6 $\pm$ 0.2) nm thick, and the Pt capping layer is (9.9 $\pm$ 0.5) nm thick, confirming good thickness uniformity across all layers.

\begin{figure}[htbp]
\includegraphics[width=0.9\columnwidth]{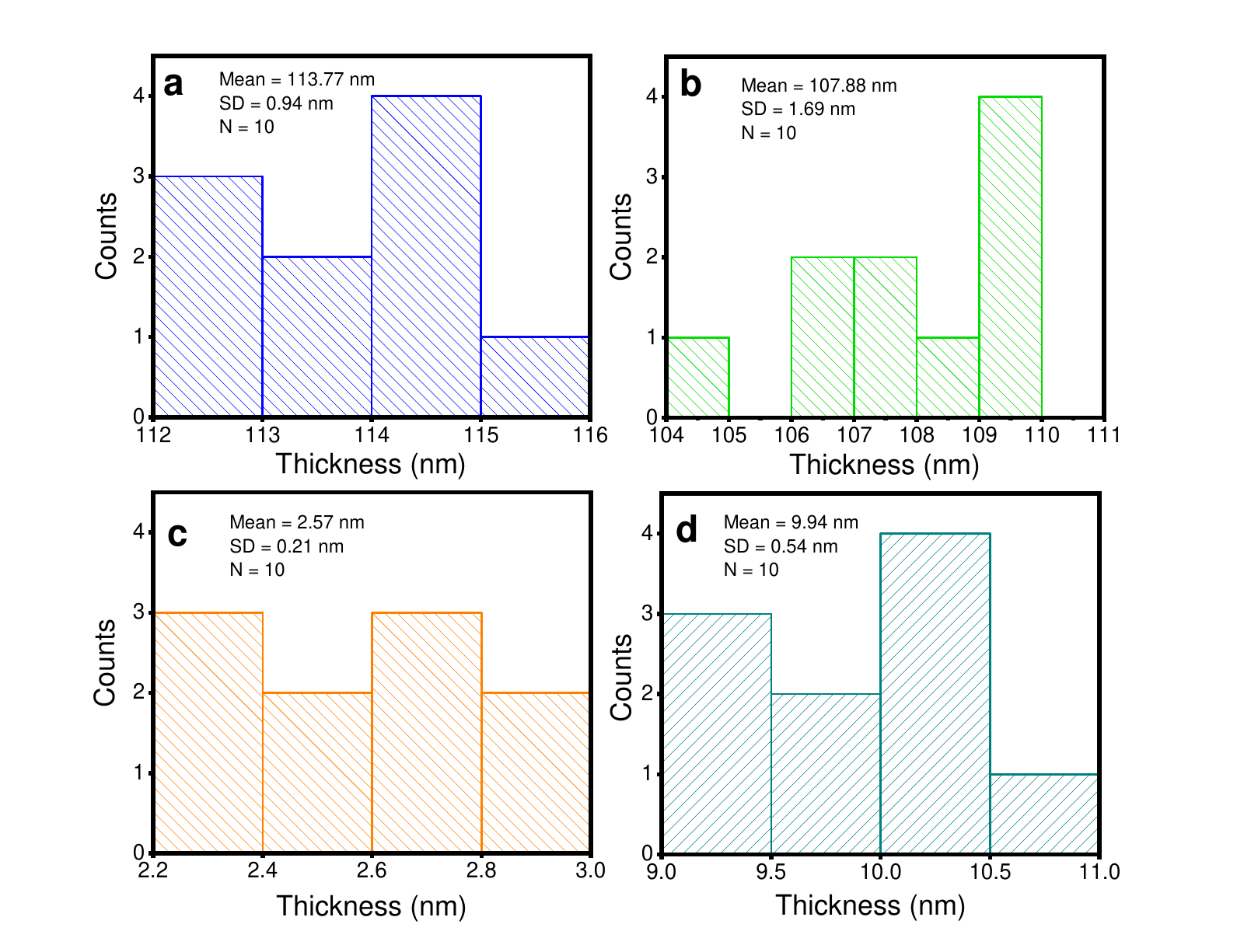}
\caption{Thickness calculations from cross-sectional TEM measurements. Histograms showing the thickness distributions for (a) Nb film without Pt capping, (b) the remaining chemically distinct Nb layer in the Pt-capped sample, (c) the native Nb oxide layer formed at the surface, and (d) the Pt capping layer deposited on Nb. The mean thickness, standard deviation (SD), and number of measurements ($N=10$) are indicated in each panel. For panel (b), the measured thickness reflects the Nb layer remaining after Pt deposition; any Nb incorporated into the interfacial Pt--Nb alloy region is not included, so the lower apparent thickness does not indicate a loss of total Nb content. Overall, the data show good thickness uniformity across all layers, with minor interfacial reaction effects only in the Pt-capped Nb sample.}
\label{FIG_SI_2}
\end{figure}

For TEM-based thickness measurements, the error bars correspond to the standard deviation obtained from multiple measurements at different regions of the cross-sectional TEM images. For each layer, $N = 10$ independent thickness measurements were performed using ImageJ. The mean thickness was calculated as
\[
\bar{x} = \frac{1}{N}\sum_{i=1}^{N}x_i,
\]
and the standard deviation was calculated as
\[
s = \sqrt{\frac{1}{N-1}\sum_{i=1}^{N}(x_i-\bar{x})^2}.
\]
The thickness values are therefore reported as $\bar{x} \pm s$. Using this approach, the Nb film without Pt capping has an average thickness of ($113.8 \pm 0.9$) nm. In the Pt-capped sample, the remaining chemically distinct Nb layer has a thickness of ($107.9 \pm 1.7$) nm. The native Nb oxide layer is ($2.6 \pm 0.2$) nm thick, and the Pt capping layer is ($9.9 \pm 0.5$) nm thick.

\clearpage

\section{Variable Energy X-ray Photoelectron Spectroscopy (VEXPS) data fitting}

Nb 3d VEXPS spectra were analyzed using a global fitting approach in which all photon energies were fitted simultaneously with a physically constrained model. This approach ensures internal  consistency of intrinsic parameters (e.g., lifetime broadening, spin–orbit splitting, and chemical shifts), while allowing extrinsic parameters (e.g., instrumental broadening and spectral weights) to vary with photon energy. The fitting model explicitly separates photon energy–independent parameters that reflect intrinsic electronic structure and local chemical environment, and photon energy–dependent parameters that capture experimental energy resolution and depth sensitivity. For the bare Nb film, the fitting model considers 5 chemical species that contain Nb, in accordance with the phase diagram of the Nb-O system \cite{naito1984phase} \ce{Nb$^0$} (bulk Nb metal), \ce{Nb^{\delta+}} (solid solution of oxygen in Nb metal), \ce{Nb^{2+}} (\ce{NbO}), \ce{Nb^{4+}} (\ce{NbO2}), and \ce{Nb^{5+}} (\ce{Nb2O5}). For the Pt/NbO$_x$/Nb films, two more species are added to model the Nb-Pt alloys, namely NbPt-1 and NbPt-2.

\subsection{Line shape models}

In the fitting, we consider the binding energies and line shapes of each species independent of photon energies, except for the instrumental-broadening that is photon-energy-dependent, but uniform across all species. The line shapes of oxide species (\ce{Nb^{2+}}, \ce{Nb^{4+}}, and \ce{Nb^{5+}}) are generally symmetric about their peaks and modeled using the Voigt profile (\texttt{lmfit.models.VoigtModel}). The \texttt{VoigtModel} is parameterized by an amplitude (integrated area) $A$, peak center $E_0$, homogeneous (Lorentzian) broadening parameter $\gamma$, and inhomogeneous (Gaussian) broadening parameter $\sigma$. Using the Faddeeva function $F(z)=e^{-z^2}\,\mathrm{erfc}(-iz)$, it is expressed as
\begin{equation}
V(E;\,E_0,\sigma,\gamma)=
\frac{A}{\sigma\sqrt{2\pi}}\,
\mathrm{Re}\!\left[F\!\left(\frac{(E-E_0)+i\gamma}{\sigma\sqrt{2}}\right)\right].
\end{equation}
In contrast, metallic species (\ce{Nb^0}, \ce{Nb^{\delta+}}, NbPt-1, and NbPt-2) exhibit intrinsically asymmetric line shapes arising from metallic screening and many-body final-state effects. These features are captured using a skewed Voigt profile (\texttt{lmfit.models.SkewedVoigtModel}), which augments the Voigt function $V(E;\,E_0,\sigma,\gamma)$ with a skewness parameter $\alpha$, yielding
\begin{equation}
V_S(E;\,E_0,\sigma,\gamma, \alpha)=V(E;\,E_0,\sigma,\gamma)\left[1+\mathrm{erf}\left(\frac{\alpha(E-E_0)}{\sigma\sqrt{2}}\right)\right].
\end{equation}

\subsection{Model parameters and constraints}

All Nb 3d components were modeled as spin–orbit doublets with a global spin-orbit splitting $\Delta E_{\mathrm{SO}} = E_{3/2} - E_{5/2}$ that is independent of photon energy and species. For each species, the chemical shift relative to \ce{Nb^0}, $\Delta E_s$, is also treated as a global parameter that is independent of photon energy. The amplitude of each species is allowed to vary independently for each photon energy, reflecting changes in probing depth and relative photoemission cross sections, while maintaining a fixed branching ratio $A_{5/2}:A_{3/2} = 3:2$. The line-shape parameters $\gamma$, $\sigma$, and $\alpha$ are subject to the following constraints:
\begin{itemize}
\item \textbf{Lorentzian broadening} ($\gamma$): The difference in core-hole lifetime may give the 3d$_{5/2}$ and 3d$_{5/2}$ branches different homogeneous broadening $\gamma$. For simplicity, all oxide species share the same pair of $\gamma_{5/2}$ and $\gamma_{3/2}$ that are independent of photon energy, while all metal species share another set. 

\item \textbf{Gaussian broadening} ($\sigma$): All collected spectra are subject to instrument broadening due to the monochromator linewidth and the photoelectron analyzer energy resolution. This broadening inherently depends on the photon energy, and can be treated as a Gaussian broadening, with a standard deviation of $\sigma_0(E)$. For each species that experiences species-dependent inhomogeneous broadening $\sigma_\mathrm{extra}(s)$, the overall inhomogeneous broadening $\sigma$ is given by 
\begin{equation}
\sigma(E,s)=\sqrt{\sigma_0^2(E)+\sigma_\mathrm{extra}^2(s)},
\end{equation}
which is shared between the 3d$_{5/2}$ and 3d$_{5/2}$ branches. Given that bulk Nb metal is the sharpest feature, we set $\sigma_\mathrm{extra}(\ce{Nb^0}) = 0$ to remove parameter redundancy. For each species other than \ce{Nb^0}, a photon-energy-independent $\sigma_\mathrm{extra}(s)$ is obtained by fitting. For convenience, we set $\sigma_\mathrm{extra}(s)$ to be identical among all oxide species. 

\item \textbf{Skewness ($\alpha$)}: The skewness parameter $\alpha$ captures many-body final-state effects arising from metallic screening and low-energy electron–hole excitations. It is independent of photon energy and set to be identical among all metallic species. The 3d$_{5/2}$ and 3d$_{5/2}$ branches also share the same $\alpha$.
\end{itemize}

\subsection{Summary of fitting parameters for bare Nb}

The fitting parameters extracted from the global Nb~3$d$ VEXPS analysis of the \ce{NbO_x}/Nb film are summarized in Table~\ref{tab:nb_global_params}, which lists the photon-energy--independent parameters, and Table~\ref{tab:energy_dependent_params}, which reports the photon-energy--dependent parameters including instrumental broadening and species fractions. Parameter uncertainties were estimated from the covariance matrix obtained from the \texttt{lmfit} least-squares optimization and are reported as one-standard-deviation ($1\sigma$) errors. Uncertainties in the species fractions were subsequently obtained through standard error propagation of the fitted peak areas, including covariance terms when available.

\begin{table}[htbp]
\centering
\caption{Photon-energy-independent fitting parameters used in the global Nb 3d VEXPS analysis for the as-grown \ce{NbO_x}/Nb film. Numbers in parentheses denote one standard error in the last digit(s), e.g., 202.215(8) represents 202.215 $\pm$ 0.008.}
\label{tab:nb_global_params}
\begin{tabular}{lcc}
\hline
\textbf{Parameter} & \textbf{Fitted Value} & \textbf{Description} \\
\hline
$E_{\mathrm{ref}}$ (eV) & 202.215(8) & Nb$^0$ 3d$_{5/2}$ reference energy \\

$\Delta E_{\mathrm{SO}}$ (eV) & 2.698(1) & Spin--orbit splitting \\

$\Delta E_{\mathrm{Nb^{\delta+}}}$ (eV) & 0.352(6) & Chemical shift of Nb$^{\delta+}$ relative to Nb$^0$ \\

$\Delta E_{\mathrm{Nb^{2+}}}$ (eV) & 2.232(10) & Chemical shift of Nb$^{2+}$ relative to Nb$^0$ \\

$\Delta E_{\mathrm{Nb^{4+}}}$ (eV) & 4.120(8) & Chemical shift of Nb$^{4+}$ relative to Nb$^0$ \\

$\Delta E_{\mathrm{Nb^{5+}}}$ (eV) & 5.608(3) & Chemical shift of Nb$^{5+}$ relative to Nb$^0$ \\

$\gamma_{5/2}^{\mathrm{metal}}$ (eV) & 0.125(1) & Metallic species 3d$_{5/2}$ lifetime width \\

$\gamma_{3/2}^{\mathrm{metal}}$ (eV) & 0.243(2) & Metallic species 3d$_{3/2}$ lifetime width \\

$\gamma_{5/2}^{\mathrm{oxide}}$ (eV) & 0.185(8) & Oxides 3d$_{5/2}$ lifetime width \\

$\gamma_{3/2}^{\mathrm{oxide}}$ (eV) & 0.252(8) & Oxides 3d$_{3/2}$ lifetime width \\

$\sigma_{\mathrm{extra}}^{\mathrm{Nb^{1+}}}$ (eV) & 0.471(4) & Additional Nb$^{\delta+}$ Gaussian width \\

$\sigma_{\mathrm{extra}}^{\mathrm{oxide}}$ (eV) & 0.412(6) & Additional oxides Gaussian width \\

$\alpha_{\mathrm{metal}}$ & 0.356(6) & Metallic species skew parameter \\
\hline
\end{tabular}
\end{table}

\begin{table}[htbp]
\centering

\setlength{\tabcolsep}{3pt}
\renewcommand{\arraystretch}{1.08}
\caption{Photon-energy ($E_{ph}$)--dependent fitting parameters for the bare Nb sample, including instrumental broadening and species fractions. Numbers in parentheses denote one standard error in the last digit(s), e.g., 0.087(2) represents 0.087 $\pm$ 0.002.}
\label{tab:energy_dependent_params}
\begin{tabular}{c c c c c c c}
\toprule
$E_{ph}$ (eV)
& $\sigma_0(E)$ (eV)
& $f_{\mathrm{Nb^0}}$
& $f_{\mathrm{Nb^{\delta+}}}$
& $f_{\mathrm{Nb^{2+}}}$
& $f_{\mathrm{Nb^{4+}}}$
& $f_{\mathrm{Nb^{5+}}}$ \\
\midrule
2000 & 0.087(2) & 0.417(3) & 0.184(3) & 0.064(2) & 0.074(2) & 0.260(2) \\
2200 & 0.103(1) & 0.482(3) & 0.168(3) & 0.059(1) & 0.064(1) & 0.227(1) \\
2500 & 0.104(2) & 0.499(3) & 0.156(3) & 0.063(2) & 0.072(2) & 0.211(2) \\
2800 & 0.117(1) & 0.580(3) & 0.132(3) & 0.052(1) & 0.073(1) & 0.164(1) \\
3000 & 0.127(1) & 0.619(4) & 0.120(3) & 0.049(2) & 0.071(1) & 0.141(1) \\
3250 & 0.087(1) & 0.639(3) & 0.123(3) & 0.060(1) & 0.054(1) & 0.124(1) \\
3500 & 0.090(1) & 0.676(3) & 0.104(3) & 0.055(1) & 0.056(1) & 0.109(1) \\
4000 & 0.100(1) & 0.725(4) & 0.091(3) & 0.048(1) & 0.047(1) & 0.089(1) \\
4500 & 0.111(2) & 0.731(5) & 0.092(4) & 0.055(2) & 0.043(2) & 0.079(2) \\
5000 & 0.166(1) & 0.767(3) & 0.091(3) & 0.050(1) & 0.033(1) & 0.059(1) \\
5500 & 0.170(1) & 0.800(4) & 0.077(4) & 0.044(2) & 0.028(1) & 0.051(1) \\
\bottomrule
\end{tabular}
\end{table}

\FloatBarrier
\subsection{Summary of fitting parameters for Pt/NbO$_x$/Nb}

The fitting parameters extracted from the global Nb~3$d$ VEXPS analysis of the Pt/NbO$_x$/Nb film are summarized in Table~\ref{tab:nbpt_global_params}, which lists the photon-energy--independent parameters, and Table~\ref{tab:nbpt_energy_dependent_params}, which reports the photon-energy--dependent parameters including instrumental broadening and species fractions. Similar as the case for \ce{NbO_x}/Nb, parameter uncertainties were estimated from the covariance matrix obtained from the \texttt{lmfit} least-squares optimization and are reported as one-standard-deviation ($1\sigma$) errors. Uncertainties in the species fractions were subsequently obtained through standard error propagation of the fitted peak areas, including covariance terms when available.

\begin{table}[htbp]
\centering
\caption{Photon-energy-independent fitting parameters used in the global Nb 3d VEXPS analysis for the Pt/NbO$_x$/Nb film. Numbers in parentheses denote one standard error in the last digit(s), e.g., 202.192(8) represents 202.192 $\pm$ 0.008.}
\label{tab:nbpt_global_params}
\begin{tabular}{lcc}
\hline
\textbf{Parameter} & \textbf{Fitted Value} & \textbf{Description} \\
\hline
$E_{\mathrm{ref}}$ (eV) & 202.192(8) & Nb$^0$ 3d$_{5/2}$ reference energy \\

$\Delta E_{\mathrm{SO}}$ (eV) & 2.751(2) & Spin--orbit splitting \\

$\Delta E_{\mathrm{Nb^{\delta+}}}$ (eV) & 0.352(6) & Chemical shift of Nb$^{\delta+}$ relative to Nb$^0$ \\

$\Delta E_{\mathrm{Nb^{2+}}}$ (eV) & 2.232(10) & Chemical shift of Nb$^{2+}$ relative to Nb$^0$ \\

$\Delta E_{\mathrm{Nb^{4+}}}$ (eV) & 3.311(14) & Chemical shift of Nb$^{4+}$ relative to Nb$^0$ \\

$\Delta E_{\mathrm{Nb^{5+}}}$ (eV) & 4.866(7) & Chemical shift of Nb$^{5+}$ relative to Nb$^0$ \\

$\Delta E_{\mathrm{NbPt-1}}$ (eV) & 0.951(9) & Chemical shift of NbPt-1 relative to Nb$^0$ \\

$\Delta E_{\mathrm{NbPt-2}}$ (eV) & 1.667(1) & Chemical shift of NbPt-2 relative to Nb$^0$ \\

$\gamma_{5/2}^{\mathrm{metal}}$ (eV) & 0.076(5) & Metallic species 3d$_{5/2}$ lifetime width \\

$\gamma_{3/2}^{\mathrm{metal}}$ (eV) & 0.150(5) & Metallic species 3d$_{3/2}$ lifetime width \\

$\gamma_{5/2}^{\mathrm{oxide}}$ (eV) & 0.125(11) & Oxides 3d$_{5/2}$ lifetime width \\

$\gamma_{3/2}^{\mathrm{oxide}}$ (eV) & 0.231(12) & Oxides 3d$_{3/2}$ lifetime width \\

$\sigma_{\mathrm{extra}}^{\mathrm{Nb^{1+}}}$ (eV) & 0.329(37) & Additional Nb$^{\delta+}$ Gaussian width \\

$\sigma_{\mathrm{extra}}^{\mathrm{oxide}}$ (eV) & 0.557(9) & Additional oxides Gaussian width \\

$\sigma_{\mathrm{extra}}^{\mathrm{Alloy\text{-}1}}$ (eV) & 0.368(15) & Additional Alloy-1 Gaussian width \\

$\sigma_{\mathrm{extra}}^{\mathrm{Alloy\text{-}2}}$ (eV) & 0.101(3) & Additional Alloy-2 Gaussian width \\

$\alpha_{\mathrm{metal}}$ & 0.238(20) & Metallic species skew parameter \\
\hline
\end{tabular}
\end{table}

\begin{table}[htbp]
\centering
\setlength{\tabcolsep}{3pt}
\renewcommand{\arraystretch}{1.08}
\caption{Photon-energy--dependent fitting parameters for the Pt/NbO$_x$/Nb film, including instrumental broadening and species fractions. Numbers in parentheses denote one standard error in the last digit(s), e.g., 0.124(4) represents 0.124 $\pm$ 0.004.}
\label{tab:nbpt_energy_dependent_params}
\begin{tabular}{c c c c c c c c c}
\toprule
Photon Energy (eV)
& $\sigma_{0}(E)$ (eV)
& $f_{\mathrm{Nb^0}}$
& $f_{\mathrm{Nb^{\delta+}}}$
& $f_{\mathrm{Nb^{2+}}}$
& $f_{\mathrm{Nb^{4+}}}$
& $f_{\mathrm{Nb^{5+}}}$
& $f_{\mathrm{Alloy\text{-}1}}$
& $f_{\mathrm{Alloy\text{-}2}}$ \\
\midrule
2000 & 0.124(4) & 0.018(2) & 0.000(5) & 0.086(7) & 0.162(3) & 0.253(2) & 0.143(7) & 0.338(6) \\
2200 & 0.130(3) & 0.029(1) & 0.000(5) & 0.100(6) & 0.161(2) & 0.222(2) & 0.158(7) & 0.331(6) \\
2500 & 0.133(4) & 0.040(2) & 0.004(6) & 0.106(7) & 0.162(3) & 0.213(2) & 0.177(8) & 0.297(6) \\
2800 & 0.143(3) & 0.056(2) & 0.005(7) & 0.136(7) & 0.157(3) & 0.151(2) & 0.202(9) & 0.292(6) \\
3000 & 0.154(3) & 0.071(3) & 0.012(8) & 0.140(6) & 0.153(3) & 0.136(2) & 0.216(9) & 0.272(6) \\
3250 & 0.106(3) & 0.095(3) & 0.018(9) & 0.137(6) & 0.133(2) & 0.111(2) & 0.237(10) & 0.270(6) \\
3500 & 0.107(3) & 0.109(3) & 0.021(9) & 0.144(6) & 0.129(2) & 0.099(2) & 0.248(10) & 0.250(6) \\
4000 & 0.115(3) & 0.149(4) & 0.041(11) & 0.143(6) & 0.113(3) & 0.075(2) & 0.261(11) & 0.217(6) \\
4500 & 0.131(3) & 0.194(5) & 0.055(12) & 0.134(6) & 0.093(3) & 0.058(3) & 0.270(11) & 0.196(5) \\
5000 & 0.155(2) & 0.225(5) & 0.068(12) & 0.142(4) & 0.079(2) & 0.050(1) & 0.269(11) & 0.167(4) \\
5500 & 0.175(3) & 0.261(6) & 0.081(13) & 0.132(4) & 0.070(2) & 0.044(2) & 0.263(11) & 0.149(4) \\
\bottomrule
\end{tabular}
\end{table}

\begin{figure}
\includegraphics[width=1.0\columnwidth]{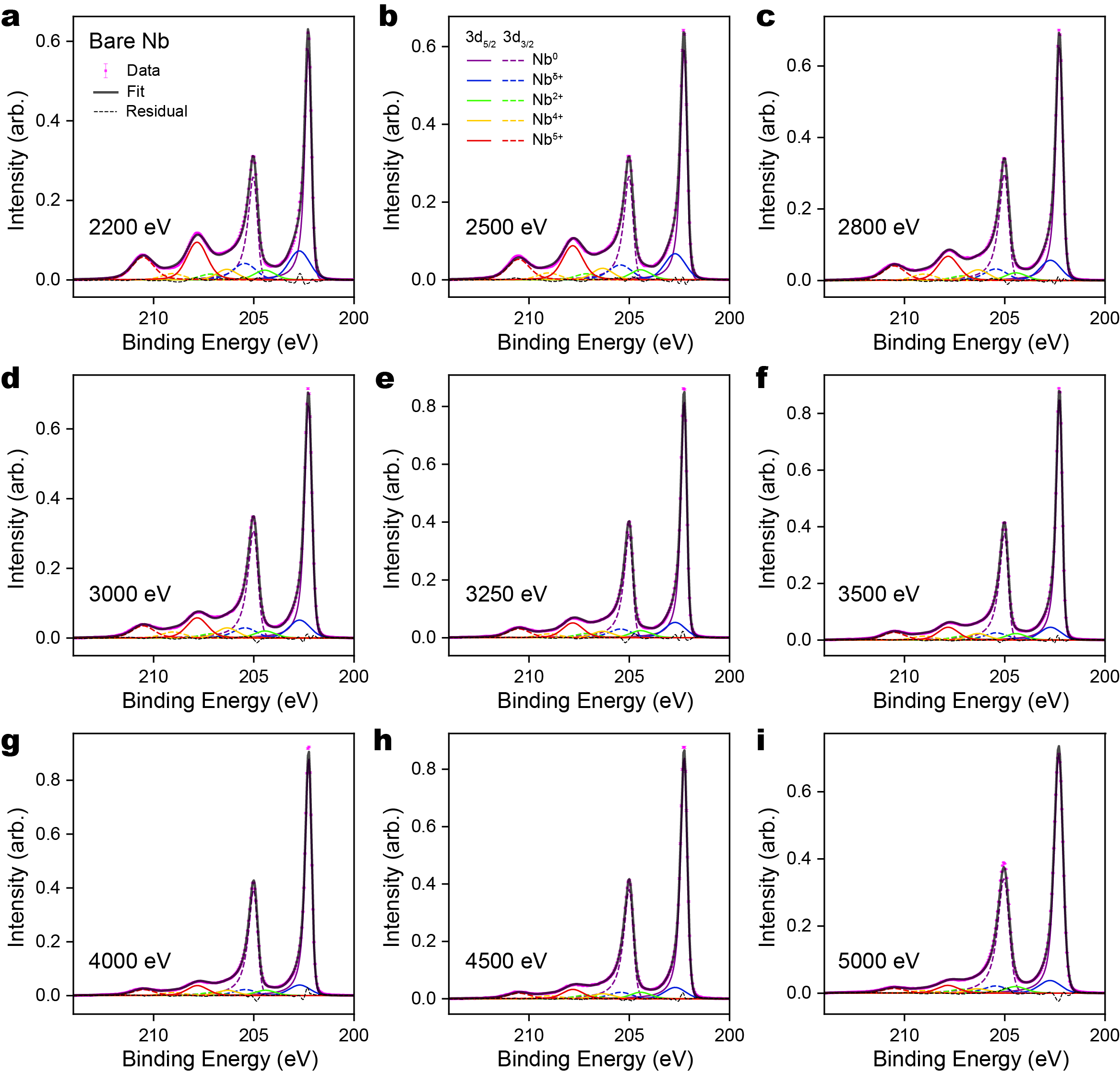}
\label{FIG SI HAXPES_bare_Nb}
\caption{(a-i) Nb 3d VEXPS spectra of NbO$_x$/Nb film, with photon energy from 2200 eV to 5000 eV.}
\end{figure}

\begin{figure}
\includegraphics[width=1.0\columnwidth]{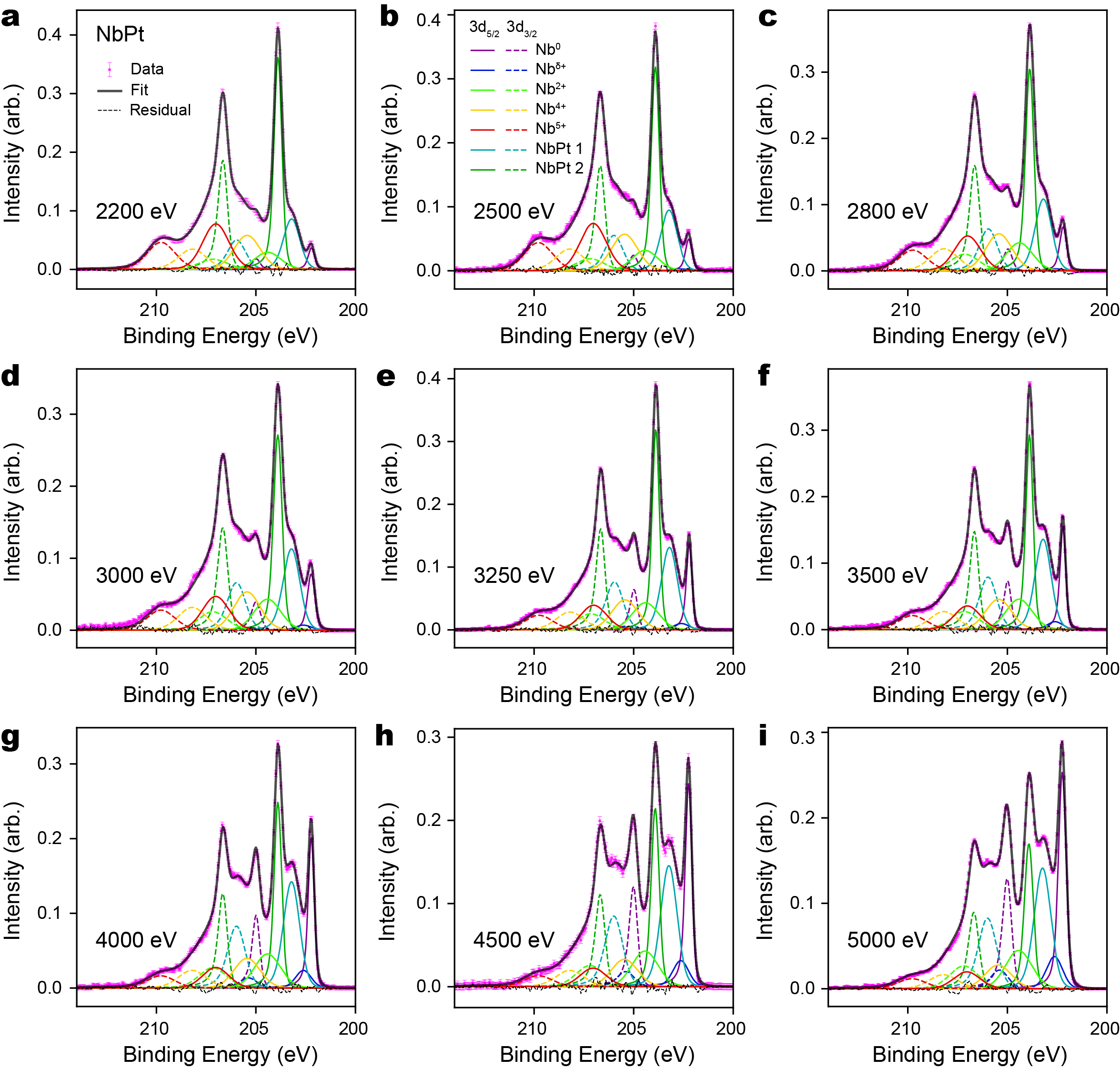}
\label{FIG SI HAXPES_NbPt}
\caption{(a-i) Nb 3d VEXPS spectra of Pt/NbO$_x$/Nb film, with photon energy ranging from 2200 eV to 5000 eV.}
\end{figure}

\clearpage

\section{Ab initio simulations}

\subsection{Structural models of NbO$_x$/Nb and Pt/NbO$_x$/Nb systems}

The Nb(111) surface was simulated using the periodic slab model containing 12 atomic planes with 4 Nb atoms per plane. The lateral supercell vectors correspond to the body-centered cubic (bcc) lattice parameter  $a_0$=3.3004 \AA. Models of partially oxidized Nb (NbO$_x$) were created by depositing 4, 8, and 12 oxygen atoms on the pure Nb(111) surface and in its vicinity, followed by the energy minimization with respect to the atomic coordinates. The resulting structures are shown in the top row in {Figure~\ref{FIG SI T-NbOx-Q}a}. 

The relaxed NbO$_x$/Nb systems were capped with one, two, and three monolayers (ML) of Pt, each monolayer containing 12 Pt atoms per lateral supercell. Representative optimized structures are shown in the bottom row in {Figure~\ref{FIG SI T-NbOx-Q}a} and in {Figure~\ref{FIG SI O-effect}a} for one and three Pt monolayers, respectively.

As a result of Pt capping, the surface oxygen atoms displace inward (i.e., away from the Pt/NbO$_x$ interface), while Nb atoms displace outward towards the Pt layers. This displacement pattern is consistent with the calculated Pt (3 ML) attachment energy of 4.5 J/m$^2$, i.e., significantly higher than typical attachment energies of metals to metals oxides \cite{hemmingson2017trends}, which points to the Pt-Nb bonds as a stabilizing factor behind the formation of Pt/NbO$_x$ interfaces.

The Nb coordination number was defined as the number of oxygen atoms within a cutoff radius ($R_c$) around the Nb atom. To define $R_c$, we calculated the total number of the Nb-O bonds depending on the distance from Nb atoms for several Pt/NbO$_x$/Nb systems (see {Figure~\ref{FIG SI T-NbOx-Q}b}). The total number of the Nb-O distances plateaus in the interval of 2.5\,\AA{} to 3.1\,\AA{}, indicating the gap between the first and second O neighboring shells near Nb atoms. We selected $R_c$ as the mid-point of this plateau ($R_c$=2.8 \AA). 

\subsection{Nb atomic charges in NbO$_x$/Nb interface}

\begin{figure}[H]
\centering
\vspace{-1.5cm}
\includegraphics[width=0.9\columnwidth]{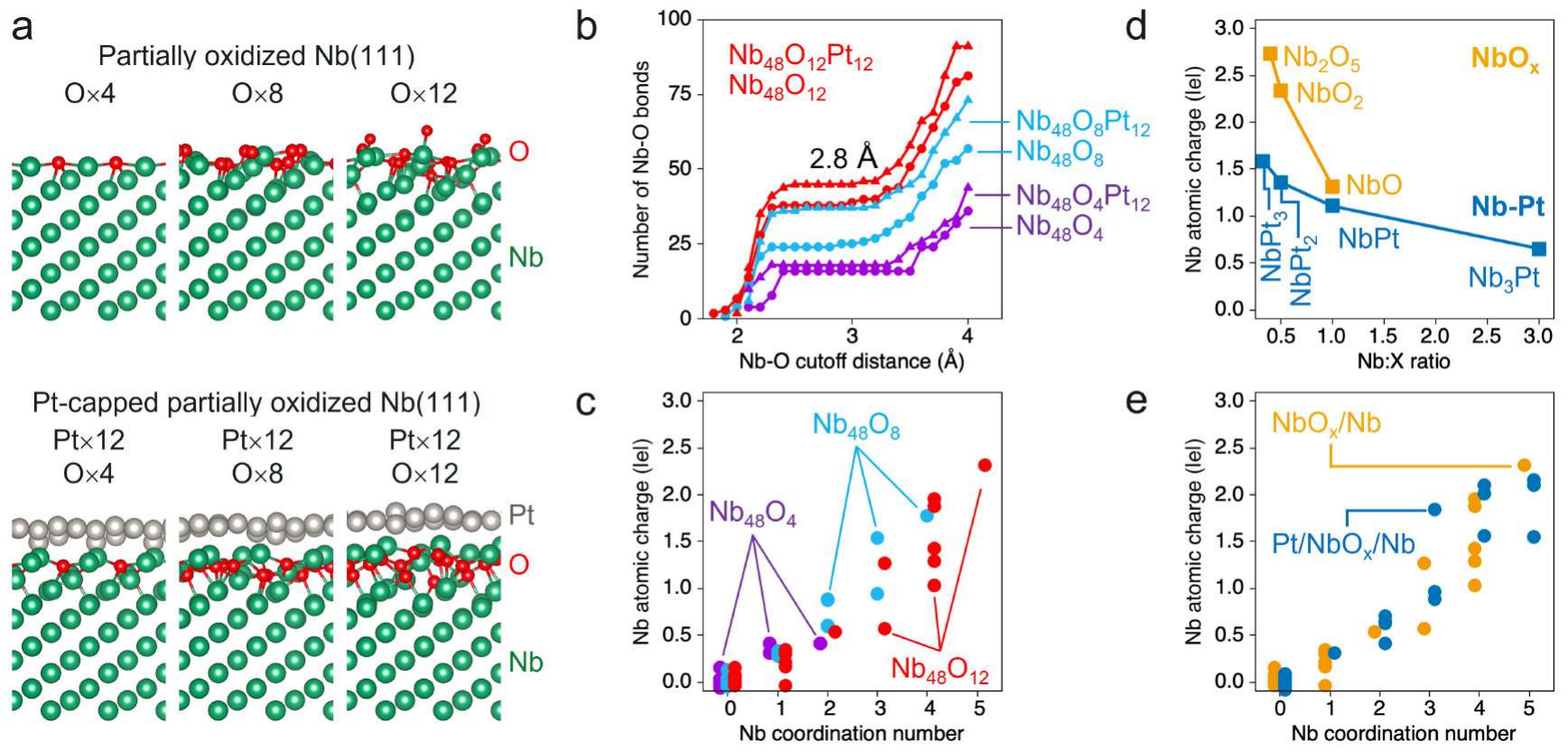}
\vspace{-1.5cm}
\caption{ 
(a) Ball-and-stick models of the Nb slab terminated with (111) surface containing 4, 8, and 12 oxygen atoms (top row) and capped with a Pt monolayer (bottom row). Nb, O, and Pt atoms are shown in green, red, and light gray, respectively.
(b) Number of the Nb-O bonds depending on the cutoff distance between Nb and O. The mid-point of the plateau (2.8 \AA) was selected as $R_c$ used to determine Nb coordination number.
(c) Nb atomic charges depending on the oxygen content and the Nb coordination number.
(d) Nb atomic charges calculated for the reference crystalline phases of Nb oxides and NbPt$_x$ intermetallics. 
(e) Comparison of the Nb atomic charges calculated without (orange) and with (dark blue) the 3 ML Pt capping layer.}

\label{FIG SI T-NbOx-Q}

\end{figure}

The Nb atomic charges were calculated using the Bader population analysis. As shown in {Figure~\ref{FIG SI T-NbOx-Q}c}, the magnitude of these charge increases with the local oxygen content represented using the Nb coordination number, reaching up +2.5 in our model oxygen-rich NbO$_x$. For comparison, Nb atomic charges in crystalline Nb oxides (NbO, NbO$_2$, Nb$_2$O$_5$) vary between approximately +1.3 and +2.7 ({Figure~\ref{FIG SI T-NbOx-Q}d}), indicating that our set of model systems covers the full range of Nb oxidation states in NbO$_x$. The trend of the Nb atomic charge in the NbPt$_x$ intermetallic compounds suggest that Pt species exhibit anionic character and, therefore, Pt capping layers in Nb-rich systems would serve as electron traps.

The effect of Pt capping on the Nb oxidation states is illustrated in {Figure~\ref{FIG SI T-NbOx-Q}e} for the case of Nb$_{48}$O$_{12}$ slab with (dark blue) and without (orange) 3 ML Pt capping layer. We observe that, upon relaxation, two of the 4-coordinated Nb atoms became 5-coordinated, which is consistent with the inward displacement of the surface oxygen species in the presence of Pt capping (compare structures with and without Pt capping in {Figure~\ref{FIG SI T-NbOx-Q}a}). We also observe that ionic charges of Nb species located next to the Pt/NbO$_x$ interface increased, indicating a possible electron transfer from NbO$_x$ to the Pt layer ({Figure~\ref{FIG SI T-NbOx-Q}e}).

\subsection{Dispersion of interstitial oxygens}

\begin{figure}[h]
\centering
\vspace{-2.0cm}
\includegraphics[width=0.9\columnwidth]{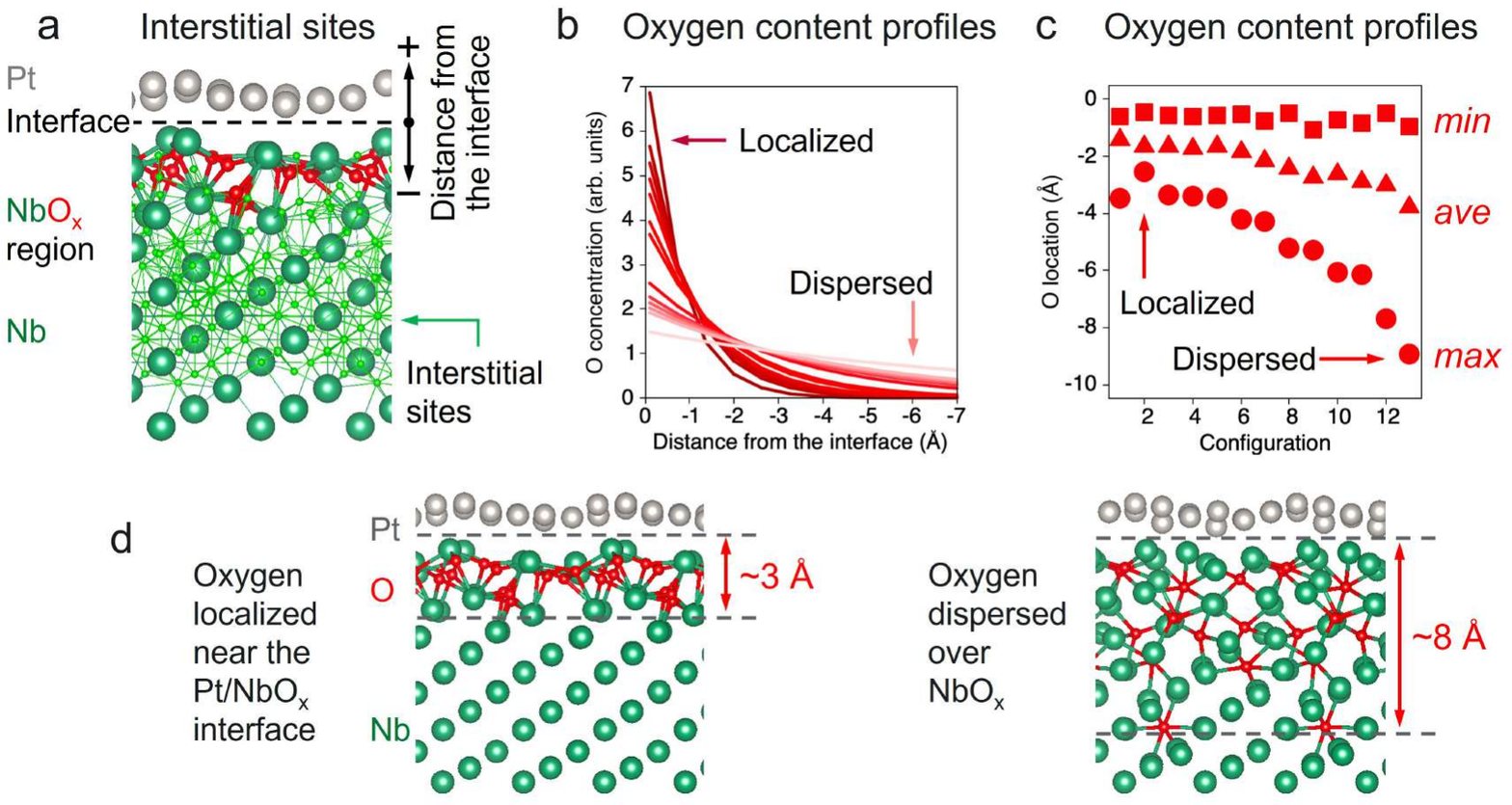}
\vspace{-1.7cm}
\caption{Simulating dispersed distributions of oxygen atoms in Pt/NbO$_x$/Nb systems. 
(a) Ball-and-stick model of Pt(1 ML)/NbO$_x$/Nb. Nb, O, and Pt atoms are shown in green, red, and light gray, respectively. The locations of possible interstitial sites are indicated with small light green spheres. 
(b) Exponential functions used to modulate the concentration profiles of the interstitial O atoms. 
(c) Characteristics of the interstitial oxygen species in fully relaxed Pt(1\,ML)/NbO$_x$/Nb systems: $min$ (squares) and $max$ (circles) correspond to the smallest and largest distances of the oxygen atoms from the Pt/NbO$_x$ interface; $ave$ (triangles) represents the geometrical center of mass of the oxygen distribution in each case. 
(d) Examples of localized (left) and dispersed (right) distributions of oxygen in Pt(1\,ML)/NbO$_x$/Nb. 
}
\label{FIG SI O-disprs}
\end{figure}

To analyze the stability of and charge distribution in Pt/NbO$_x$/Nb depending on the spatial distribution of oxygen atoms, we generated configurations of Pt/NbO$_x$/Nb systems (Pt coverage of 0, 1, and 3 ML) in which the width of the NbO$_x$ layer was progressively increased, while the total composition of the NbO$_x$/Nb systems remained fixed (Nb$_{48}$O$_{12}$). To define the initial locations of the oxygen atoms across NbO$_x$/Nb, we constructed the graph of the nearest Nb-Nb neighbors within the cutoff distance of 3.5 \AA\ and identified second-nearest Nb-Nb pairs. The mid-points between all second-nearest neighbors were selected as candidates for interstitial O sites. Then, sites that are closer than 1.4 \AA\ to NbO$_x$/Nb atoms were eliminated, while the sites already occupied by oxygen atoms in NbO$_x$/Nb systems were included into the list. An example of the resulting set of interstitial sites is shown in { Figure~\ref{FIG SI O-disprs}a}.

Then, the maximum targeted width of the NbO$_x$ layer (8 \AA) was split into 12 intervals, 0.67 \AA\ each, and the occupancy of oxygen in each bin was modulated with exponentially decaying functions (see { Figure~\ref{FIG SI O-disprs}b}) under the condition that the total number of oxygen atom remains constant; specific sites occupied by O atoms in each bin were selected at random. 13 configurations generated using this approach were fully relaxed for each Pt coverage (0, 1, and 3 ML). The distribution of oxygen atoms corresponding to the Pt coverage of 1 ML is characterized in { Figure~\ref{FIG SI O-disprs}c} in terms of the closest and furthest O locations to the Pt/NbO$_x$ interface ($min$ and $max$, respectively). The center of mass of the O atoms in NbO$_x$ is shown as $ave$. Examples of the least dispersed (localized) and the most dispersed configurations are shown in { Figure~\ref{FIG SI O-disprs}d})

\begin{figure}[htbp]
\includegraphics[width=0.9\columnwidth]{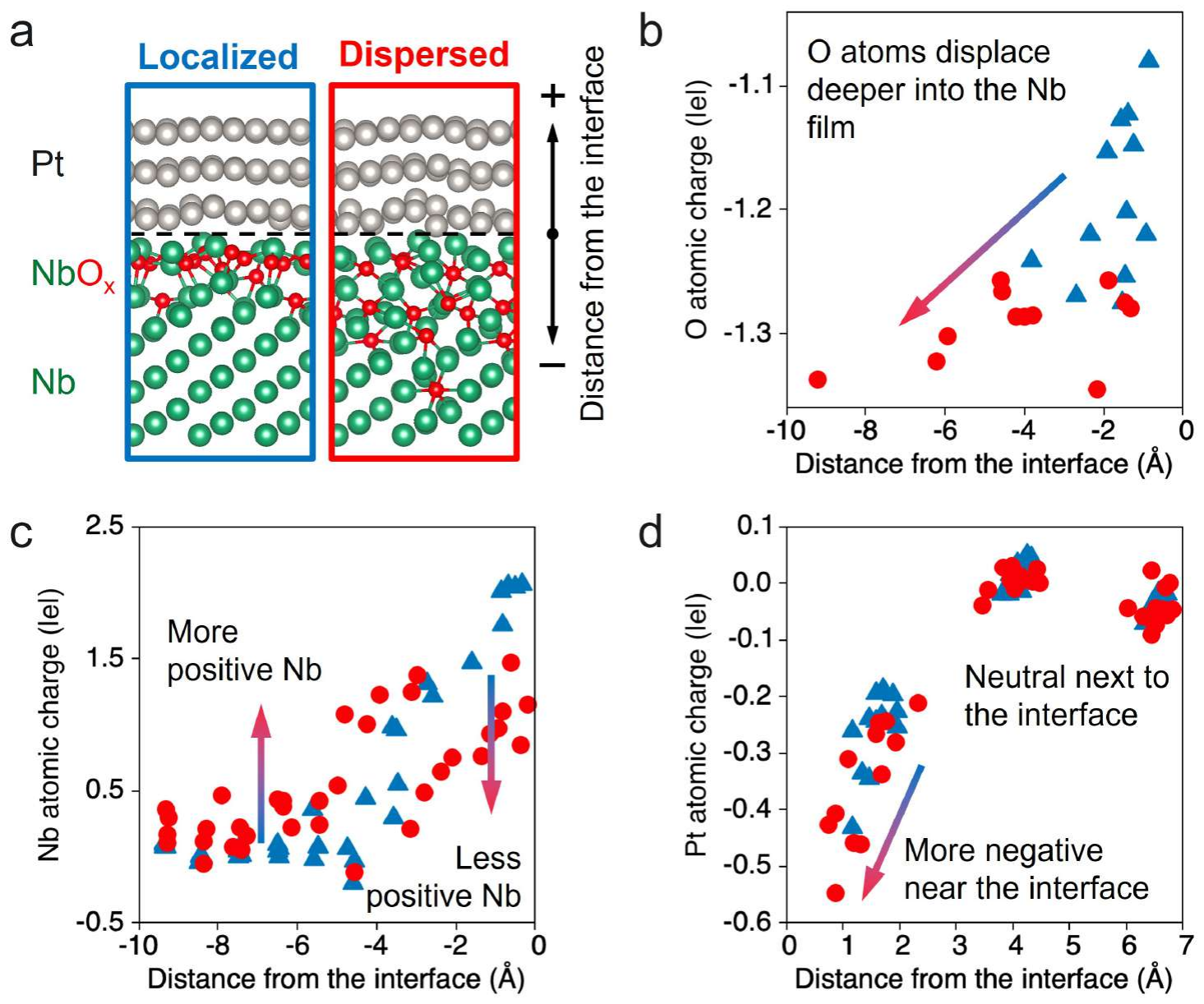}
\caption{
(a) Ball-and-stick representations of the localized and dispersed distributions of oxygen species in the Pt (3 ML)/NbOx/Nb systems. Nb, O, and Pt atoms are shown in green, red, and light gray, respectively.
(b-d) Changes of the O, Nb, and Pt atomic charges induced by the transition from the localized (blue) to the dispersed (red) oxygen distribution.
}
\label{FIG SI O-effect}
\end{figure}

The effect of the oxygen dispersion on the local atomic structure of NbO$_x$ is illustrated in {Figure~\ref{FIG SI O-effect}a}, where we compare the O, Nb, and Pt atomic charges for the localized and the most dispersed O configurations (1 and 13 in Figure~\ref{FIG SI O-disprs}, respectively, albeit with Pt coverage of 3 ML). The larger corrugation of the interfacial Pt plane in the limit of high dispersion indicates a stronger interaction of the Pt film with the NbO$_x$ substrate. 

The qualitative changes of the charge distribution in { Figure~\ref{FIG SI O-effect}b-d} are indicated with color-coded arrows with the tail (blue) corresponding to the localized configuration and the arrowhead (red) corresponding to the dispersed configuration. Negative distances from the interface in {Figure~\ref{FIG SI O-effect}} and elsewhere corresponds to the inward direction into Nb bulk, while positive distances correspond to the outward direction into the Pt layer.

The broadening of the NbO$_x$ layer means that O atoms appear at larger distances from the Pt/NbO$_x$ interface. Since the local ratio of Nb:O increases, the O atoms tend to pull a higher amount of the electron density from their Nb neighbors, resulting in a more negative charge ({Figure~\ref{FIG SI O-effect}b}). Accordingly, Nb atoms next to the Pt/NbO$_x$ interface, become less positive, while Nb atoms deeper in the Nb bulk become more positive {Figure~\ref{FIG SI O-effect}c}. Finally, Pt atoms forming the Pt/NbO$_x$ interface become slightly more negative, indicating that oxygen depletion near the interface promotes charge transfer to the second most electronegative species (Pt). For comparison, Pt atoms in the second and third atomic planes remain neutral. 

\subsection{Nb-Pt intermixing}

\begin{figure}[htbp]
\vspace{-3cm}
\includegraphics[width=0.9\columnwidth]{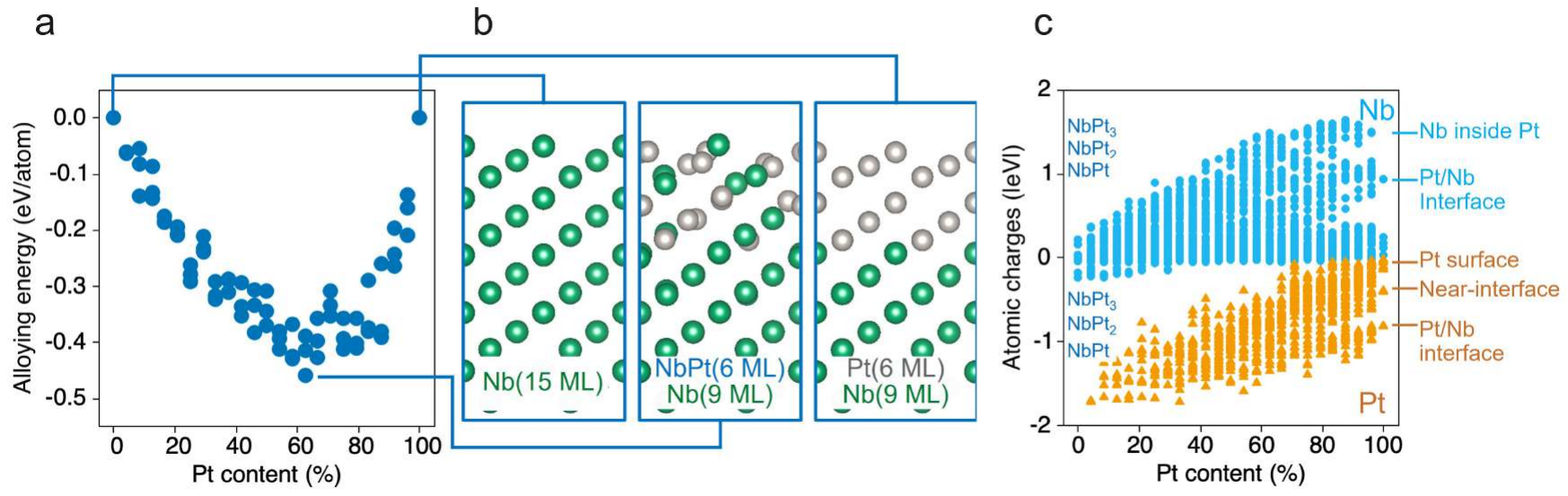}
\vspace{-2.5 cm}

\caption{
(a) Nb-Pt alloying energy calculated for the intermixing window of 6 planes (out of 15 planes total) near the Nb(111) surface.
(b) Selected configurations: Nb slab terminated with the (111) surface (left), 6-planes thick Pt layer in the bcc lattice structure (right) on the Nb slab, the most stable configuration found in these calculations (center). Nb and Pt atoms are shown in green and light gray, respectively.
(c) Nb and Pt atomic charges for each considered Nb-Pt configuration, plotted vs Pt content in the intermixing window. Dashed lines indicate the Nb and Pt charges in crystalline NbPt$_x$ bulk phases for comparison.
}
\label{FIG SI Nb-Pt-mix}
\end{figure}

The effects of Nb-Pt intermixing were simulated using a slab model comprised of 15 atomic planes (4 atoms per plane) of Nb terminated with (111) surface. The outermost 6 planes of this slab were used as intermixing window where the concentration of Pt was varied from 0 \% to 100 \% (from 0 atoms to 24 atoms) in the increments of 1 Pt atom. The locations of substitutional Pt atoms in the bcc Nb lattice was selected at random. Three configurations of the substitutional Pt atoms were considered for each case except for the end-points of 0 \% and 100 \%. For each configuration, the applying energy $E_a$ (see {Figure~\ref{FIG SI Nb-Pt-mix}a}) was calculated as

$$E_a = \frac{1}{N_{\text{Pt}} + N_{\text{Nb}}} \cdot \left [ E_{\text{Nb-Pt}} - N_{\text{Pt}}*E_{\text{Pt}} - N_{\text{Nb}}*E_{\text{Nb}} + f(N_{\text{Pt}}) \right],$$
where $E_{\text{Nb-Pt}}$ is the total energy of the slab containing intermixed Nb-Pt layer; $N_{\text{Pt}}$ and $N_{\text{Nb}}$ are the numbers of Pt and Nb atoms in the slab, respectively; $E_{\text{Pt}}$ and $E_{\text{Nb}}$ are the bulk energies of Pt and Nb per atoms; $f(N_{\text{Pt}})$ is a linear function defined so that to place the value of $E_a$ for the end-points of 0 \% and 100\% are zero ({Figure~\ref{FIG SI Nb-Pt-mix}a}). Atomic configurations of these end-points and the lowest-energy configuration are shown in {Figure~\ref{FIG SI Nb-Pt-mix}b}. 

Atomic charges corresponding the Nb and Pt atoms for all Nb-Pt configurations are shown in {Figure~\ref{FIG SI Nb-Pt-mix}c}. As the concentration of Pt atoms in the intermixing window increases, the charges of Nb atoms become more positive and the charges of Pt atoms -- less negative, in line with the changes in the crystalline NbPt$_x$ phases. In the limit of high Pt content, the Pt atoms at the Pt/Nb interface retain the charge of approximately --1e. However, the charge magnitude decreases rapidly with the distance from the interface. In particular, Pt atoms next to the interfacial plane have the charge of only --0.3e, while the Pt surface is neutral. Similarly fast decay of the ionic charge with the distance from the interface is observed on the Nb side.

\clearpage

\section{Extended X-ray Absorption Fine Structure (EXAFS) Spectroscopy study and Analysis}

To determine the local atomic structure of the Pt overlayer and underlying Nb, Pt $L_3$-edge and Nb $K$-edge EXAFS spectra were analyzed using a series of candidate structural models. The Pt $L_3$-edge data were tested against bulk fcc Pt, NbPt$_2$, NbPt$_3$, and equiatomic NbPt models to identify the local coordination environment of Pt at the buried interface. The Nb $K$-edge spectra were fit using a bulk bcc Nb model to assess whether the underlying Nb lattice remained structurally intact after Pt deposition and annealing. The corresponding fitting results are summarized in Tables V--VII, and the comparison between experimental data and model fits is shown in Figures S9--S11.

\begin{figure}[htbp]

\includegraphics[width=0.9\columnwidth]{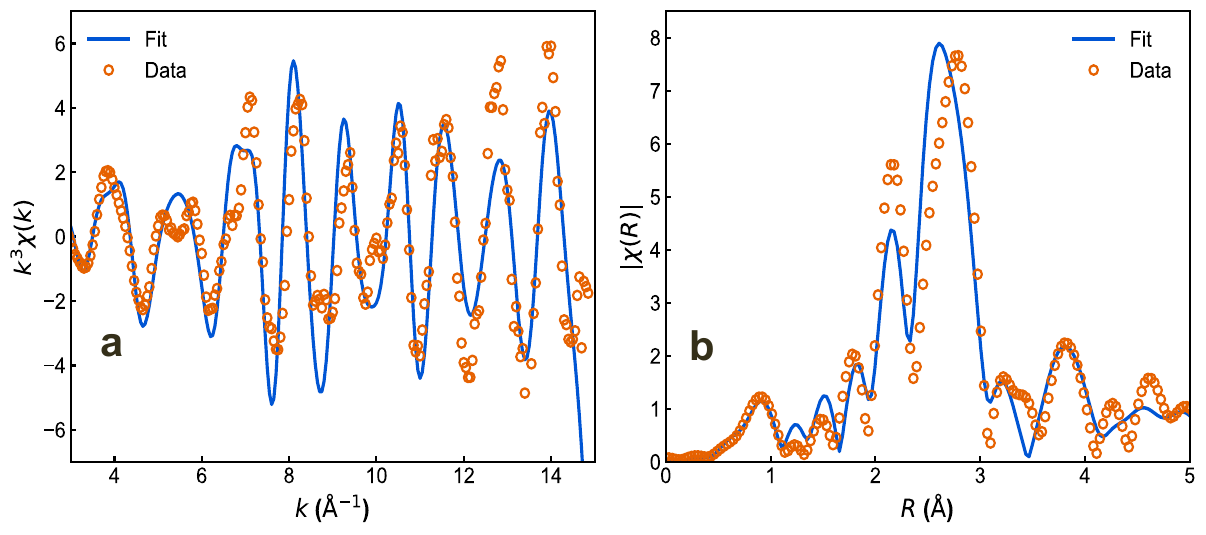}
\label{FIG SI 3.}
\caption{Failure of the metallic fcc Pt model to describe the Pt/NbO$_x$/Nb interfacial structure. 
(a) k$^3$-weighted Pt L$_3$-edge EXAFS oscillations, $\chi(k)\cdot k^3$, for the annealed Pt/NbO$_x$/Nb thin film (symbols) together with the best-fit simulation using a bulk fcc Pt structural model (solid line). Although the fitting protocol was validated against a Pt foil reference, the metallic Pt model fails to reproduce the experimental oscillation amplitude and phase across the full $k$-range (3.0~\AA$^{-1}$ to 14.5~\AA$^{-1}$), yielding large systematic residuals. 
(b) Magnitude of the Fourier-transformed EXAFS signal $|\chi(R)|$ with fitting performed over 1.6~\AA{} to 5~\AA{}. The model exhibits pronounced discrepancies in both first- and higher-shell peak positions and amplitudes. The resulting elevated reduced $\chi^2$ and R-factor demonstrate that the Pt overlayer does not retain bulk fcc coordination, providing direct spectroscopic evidence of interfacial alloy formation.
}
\end{figure}

\begin{figure}
\includegraphics[width=0.9\columnwidth]{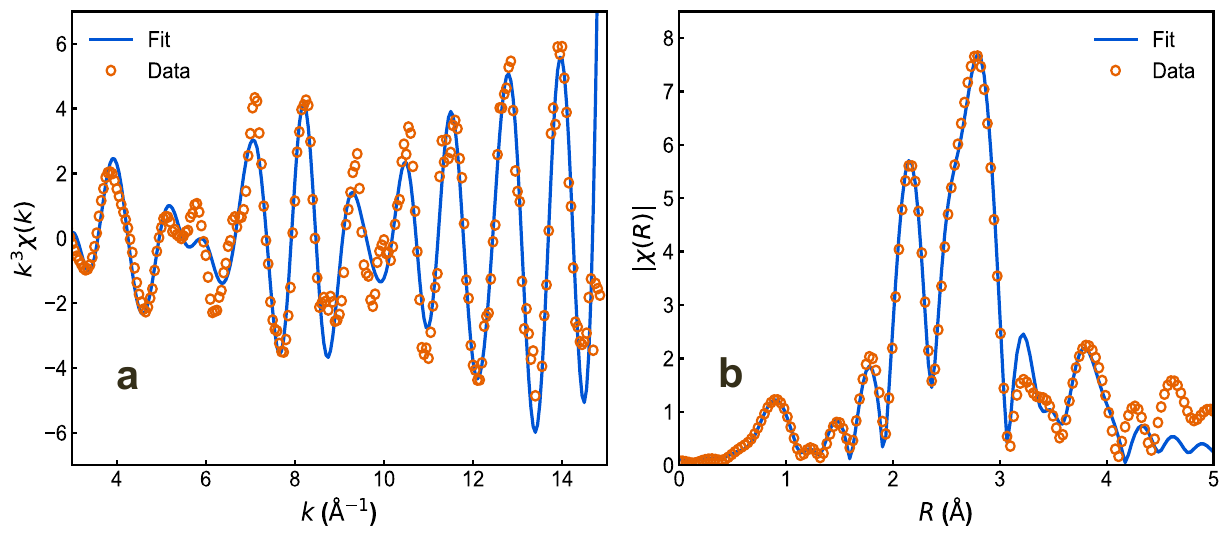}
\label{FIG SI 3.}
\caption{NbPt$_2$ structural model provides an inadequate description of the Pt local environment. 
(a) k$^3$-weighted Pt L$_3$-edge EXAFS oscillations, $\chi(k)\cdot k^3$, for the annealed Pt/NbO$_x$/Nb thin film (symbols) fitted using an NbPt$_2$ structural model (solid line). Partial agreement is observed in the first-shell region; however, clear phase and amplitude mismatches persist at higher $k$ values, indicating incomplete reproduction of the experimental oscillations. 
(b) Magnitude of the Fourier-transformed EXAFS signal $|\chi(R)|$ with fitting performed over 1.6~\AA{} to 4.1~\AA{}. The NbPt$_2$ model does not accurately capture higher-shell peak positions and amplitudes, resulting in elevated residuals and a substantially larger reduced $\chi^2$ compared to the NbPt$_3$ model. These observations demonstrate that NbPt$_2$ cannot represent the dominant local coordination environment of Pt at the interface and may contribute only as a minor secondary phase.
}
\end{figure}

\begin{figure}
\includegraphics[width=0.9\columnwidth]{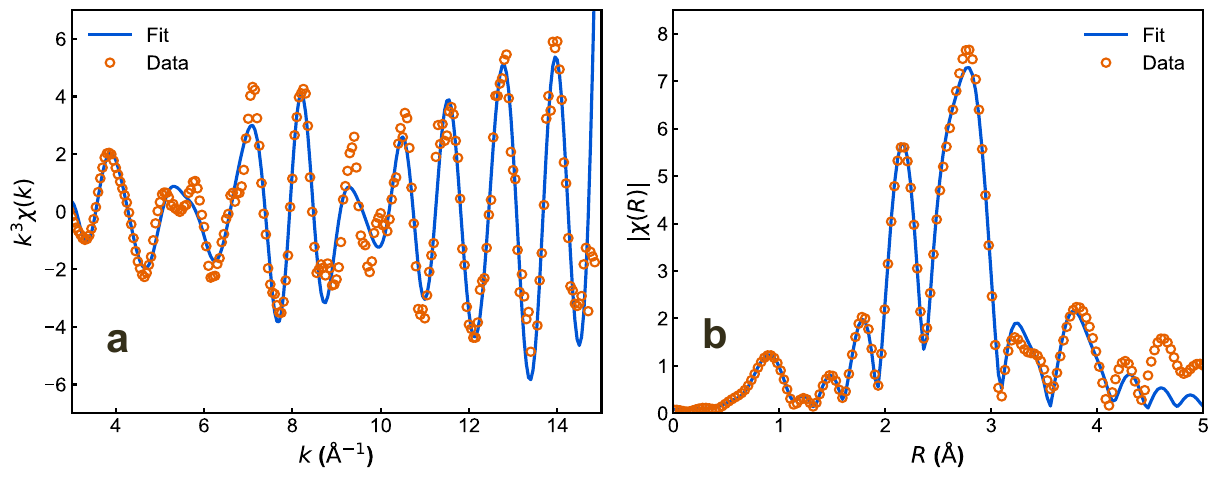}
\label{FIG SI 3.}
\caption{Equiatomic NbPt model yields unphysical structural parameters and provides an inadequate description of the Pt local coordination. 
(a) k$^3$-weighted Pt L$_3$-edge EXAFS oscillations, $\chi(k)\cdot k^3$, for the annealed Pt/NbO$_x$/Nb thin film (symbols) fitted using a NbPt structural model (solid line). Although partial agreement is observed in limited $k$ regions, substantial phase and amplitude deviations persist across the full $k$-range (3.0~\AA$^{-1}$ to 14.5~\AA$^{-1}$), indicating an inadequate structural description. 
(b) Magnitude of the Fourier-transformed EXAFS signal $|\chi(R)|$ with fitting performed over 1.6~\AA{} to 4.1~\AA{}. The NbPt model exhibits pronounced discrepancies in both first- and higher-shell peaks. Refinement yields chemically implausible parameters, including an unrealistically short Pt–Nb bond length (approximately 1.99~\AA) and unstable Debye–Waller factors, demonstrating that equiatomic NbPt cannot represent the dominant local coordination environment at the Pt/Nb interface.
}
\end{figure}

\FloatBarrier

\clearpage

\begin{table}[htbp]
\caption{Key first-shell fitting parameters (Pt absorber).}
\label{tab:SI7}
\centering
\begin{tabular}{l l c c c}
\hline
Model & Scattering path & $N$ & $R$ (\AA) & $\sigma^2$ (\AA$^2$) \\
\hline
NbPt$_3$ & Pt--Pt & 5 & 2.756 & 0.0039 \\
         & Pt--Nb & 4 & 2.781 & 0.0090 \\
NbPt$_2$ & Pt--Pt & 3--4 & 2.77--2.86 & 0.0006--0.0083 \\
NbPt     & Pt--Nb & 4 & 1.99 & 0.0286 \\
\hline
\end{tabular}
\end{table}

The best-fit \ce{NbPt3} model yields bond lengths of $(2.756 \pm 0.001)$ \AA\ for Pt--Pt and $(2.781 \pm 0.003)$ \AA\ for Pt--Nb; the uncertainties are one standard deviation from the Artemis covariance matrix, and comparison-model values are reported for reference.

\begin{table*}[htbp]
\caption{Summary of Pt $L_3$-edge EXAFS fitting results for Nb--Pt alloy models.}
\label{tab:SI8}
\centering
\begin{tabular}{l c c c c c c l}
\hline
Model (mp-ID) & $N_{\mathrm{ind}}$ & $N_{\mathrm{var}}$ & $\chi^2_{\mathrm{red}}$ & R-factor & $S_0^2$ & $\Delta E_0$ (eV) & Structural assessment \\
\hline
NbPt$_3$ (mp-12700)  & 29.52 & 14 & 2.38 & 0.0056 & 0.70 & 8.00 & Excellent fit, physically reasonable \\
NbPt$_2$ (mp-11514)  & 29.52 & 14 & 8.21 & 0.0193 & 0.70 & 8.00 & Poor agreement, higher residuals \\
NbPt (mp-999376)     & 29.52 & 17 & 4.32 & 0.0082 & 0.97 & 5.62 & Unphysical $\Delta R$, unstable $\sigma^2$ \\
\hline
\end{tabular}
\end{table*}

\begin{figure}
\includegraphics[width=0.9\columnwidth]{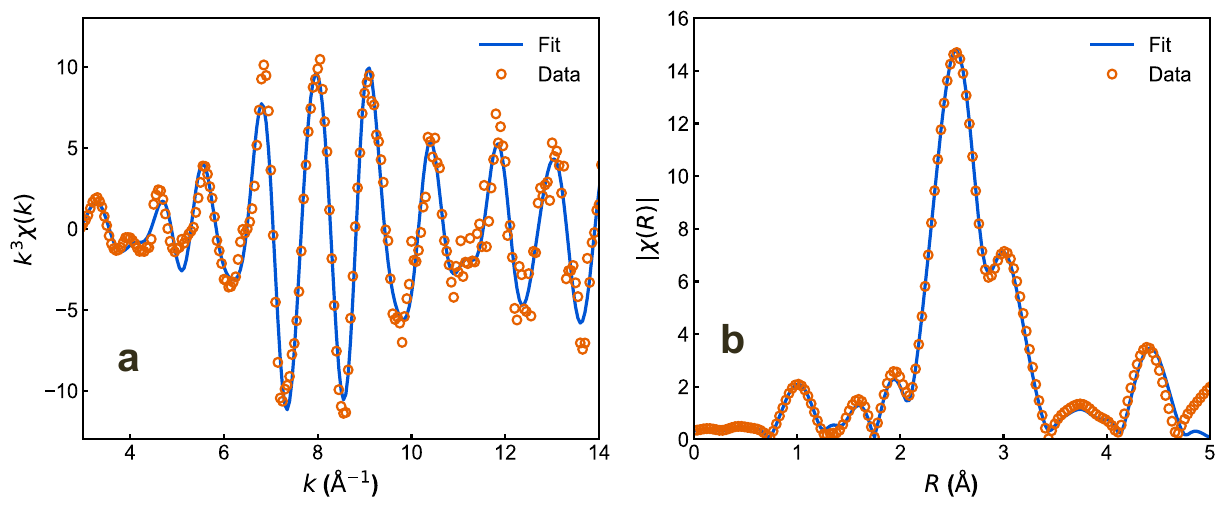}
\label{FIG SI 3.}
\caption{Nb K-edge EXAFS confirms preservation of the bulk bcc Nb structure beneath the Pt overlayer. 
(a) k$^3$-weighted Nb K-edge EXAFS oscillations, $\chi(k)\cdot k^3$, for the annealed Pt/NbO$_x$/Nb thin film (symbols) together with the best-fit simulation using a bulk bcc Nb model (solid line). The high-quality agreement across the measured $k$-range demonstrates that the underlying Nb lattice remains structurally intact following Pt deposition and annealing. 
(b) Magnitude of the Fourier-transformed EXAFS signal $|\chi(R)|$ with fitting performed over 1.6~\AA{} to 5~\AA{}, showing that both the first- and higher-shell coordination peaks are well reproduced. The excellent fit quality (low reduced $\chi^2$ and R-factor) confirms that interfacial alloying is confined to the near-surface Pt/Nb region, while the bulk Nb maintains its original bcc structure, consistent with STEM observations and superconducting transport measurements.
}
\end{figure}

\FloatBarrier

\begin{table}[htbp]
\caption{Nb K-edge EXAFS fitting results.}
\label{tab:SI10}
\centering
\begin{tabular}{l c c l}
\hline
Sample & $\chi^2_{\mathrm{red}}$ & R-factor & Structural assessment \\
\hline
Nb film & 1.36 & 0.0034 & Bulk-like bcc Nb \\
\hline
\end{tabular}
\end{table}

For EXAFS analysis, the spectra were processed in Athena and fitted in Artemis within the Demeter/IFEFFIT framework. The uncertainties in the fitted structural parameters were obtained from the covariance matrix of the final least-squares fit in Artemis. The reported bond-length uncertainties correspond to one standard deviation returned by the fit. These values represent fitting uncertainties associated with the optimized EXAFS model parameters and should not be interpreted as absolute experimental errors. The reduced $\chi^2$ and $R$-factor values indicate that the fitted parameters are well constrained by the data.

\clearpage

%

\clearpage
\section*{References}
\bibliography{apssamp}
\end{document}